\documentclass[12pt,a4paper]{article}

\usepackage{ifthen,placeins} 
\newboolean{pdflatex}
\setboolean{pdflatex}{true} 

\newboolean{articletitles}
\setboolean{articletitles}{true} 

\newboolean{uprightparticles}
\setboolean{uprightparticles}{false} 

\newboolean{inbibliography}
\setboolean{inbibliography}{false} 

\usepackage[top=1in, bottom=1.25in, left=1in, right=1in]{geometry}

\usepackage{microtype}
\usepackage{lineno}  
\usepackage{xspace} 
\usepackage{caption} 

\usepackage{graphicx}  
\usepackage{color}
\usepackage{colortbl}
\graphicspath{{./figs/}} 

\usepackage{amsmath} 
\usepackage{amssymb}
\usepackage{amsfonts}
\usepackage{upgreek} 

\newcommand*\patchAmsMathEnvironmentForLineno[1]{%
\expandafter\let\csname old#1\expandafter\endcsname\csname #1\endcsname
\expandafter\let\csname oldend#1\expandafter\endcsname\csname
end#1\endcsname
 \renewenvironment{#1}%
   {\linenomath\csname old#1\endcsname}%
   {\csname oldend#1\endcsname\endlinenomath}%
}
\newcommand*\patchBothAmsMathEnvironmentsForLineno[1]{%
  \patchAmsMathEnvironmentForLineno{#1}%
  \patchAmsMathEnvironmentForLineno{#1*}%
}
\AtBeginDocument{%
\patchBothAmsMathEnvironmentsForLineno{equation}%
\patchBothAmsMathEnvironmentsForLineno{align}%
\patchBothAmsMathEnvironmentsForLineno{flalign}%
\patchBothAmsMathEnvironmentsForLineno{alignat}%
\patchBothAmsMathEnvironmentsForLineno{gather}%
\patchBothAmsMathEnvironmentsForLineno{multline}%
\patchBothAmsMathEnvironmentsForLineno{eqnarray}%
}

\usepackage{hyperref}    
\usepackage[all]{hypcap} 

\newcommand{\Pom}{\mathbb{P}}
\renewcommand{\d}{{\mathrm d}}
\newcommand{\avg}[1]{\left< #1 \right>} 

\def\MagUp {\mbox{\em Mag\kern -0.05em Up}\xspace}

\ifthenelse{\boolean{uprightparticles}}%
{

 \def\Peta        {\ensuremath{\upeta}\xspace}

 \def\Pmu         {\ensuremath{\upmu}\xspace}

 \def\PDelta      {\ensuremath{\Delta}\xspace}                 
 \def\PXi      {\ensuremath{\Xi}\xspace}                 
 \def\PLambda      {\ensuremath{\Lambda}\xspace}                 
 \def\PSigma      {\ensuremath{\Sigma}\xspace}                 
 \def\POmega      {\ensuremath{\Omega}\xspace}                 
 \def\PUpsilon      {\ensuremath{\Upsilon}\xspace}                 
 
 \def\PB      {\ensuremath{\mathrm{B}}\xspace}                 
                  
 \def\PD      {\ensuremath{\mathrm{D}}\xspace}

 \def\PK      {\ensuremath{\mathrm{K}}\xspace}

 \def\PW      {\ensuremath{\mathrm{W}}\xspace}

 \def\PZ      {\ensuremath{\mathrm{Z}}\xspace}

 \def\Pi      {\ensuremath{\mathrm{i}}\xspace}

}
{

 \def\Peta        {\ensuremath{\eta}\xspace}

 \def\Pmu         {\ensuremath{\mu}\xspace}

 \mathchardef\PDelta="7101
 \mathchardef\PXi="7104
 \mathchardef\PLambda="7103
 \mathchardef\PSigma="7106
 \mathchardef\POmega="710A
 \mathchardef\PUpsilon="7107
                  
 \def\PB      {\ensuremath{B}\xspace}                 
                  
 \def\PD      {\ensuremath{D}\xspace}

 \def\PK      {\ensuremath{K}\xspace}

 \def\PW      {\ensuremath{W}\xspace}

 \def\PZ      {\ensuremath{Z}\xspace}

 \def\Pi      {\ensuremath{i}\xspace}

}

\makeatletter
\ifcase \@ptsize \relax
  \newcommand{\miniscule}{\@setfontsize\miniscule{4}{5}}
\or
  \newcommand{\miniscule}{\@setfontsize\miniscule{5}{6}}
\or
  \newcommand{\miniscule}{\@setfontsize\miniscule{5}{6}}
\fi
\makeatother

\DeclareRobustCommand{\optbar}[1]{\shortstack{{\miniscule (\rule[.5ex]{1.25em}{.18mm})}
  \\ [-.7ex] $#1$}}

\def\muon       {{\ensuremath{\Pmu}}\xspace}

\def\Wp     {{\ensuremath{\PW^+}}\xspace}
\def\Wm     {{\ensuremath{\PW^-}}\xspace}
\def\Wpm    {{\ensuremath{\PW^\pm}}\xspace}
\def\Z      {{\ensuremath{\PZ}}\xspace}

\def\rwpm {{\ensuremath{R_{\Wpm}}}\xspace}

\def\rwpz {{\ensuremath{R_{\Wp\PZ}}}\xspace}
\def\rwmz {{\ensuremath{R_{\Wm\PZ}}}\xspace}

\def\antikt     {\ensuremath{\text{anti-}k_{\mathrm{T}}}\xspace}

\def\fastjet    {\mbox{\textsc{FastJet}}\xspace}

  \def\Kbar    {{\kern 0.2em\overline{\kern -0.2em \PK}{}}\xspace}

\def\KorKbar    {\kern 0.18em\optbar{\kern -0.18em K}{}\xspace}

  \def\Dbar    {{\kern 0.2em\overline{\kern -0.2em \PD}{}}\xspace}

\def\DorDbar    {\kern 0.18em\optbar{\kern -0.18em D}{}\xspace}

\def\Bbar    {{\ensuremath{\kern 0.18em\overline{\kern -0.18em \PB}{}}}\xspace}

\def\BorBbar    {\kern 0.18em\optbar{\kern -0.18em B}{}\xspace}

  \def\Y#1S{\ensuremath{\PUpsilon{(#1S)}}\xspace}

\def\Lbar        {{\ensuremath{\kern 0.1em\overline{\kern -0.1em\PLambda}}}\xspace}
\def\LorLbar    {\kern 0.18em\optbar{\kern -0.18em \PLambda}{}\xspace}

\def\AT#1     {\ensuremath{A_{\mathrm{T}}^{#1}}\xspace}           

\def\C#1      {\ensuremath{\mathcal{C}_{#1}}\xspace}                       
\def\Cp#1     {\ensuremath{\mathcal{C}_{#1}^{'}}\xspace}                    
\def\Ceff#1   {\ensuremath{\mathcal{C}_{#1}^{\mathrm{(eff)}}}\xspace}        
\def\Cpeff#1  {\ensuremath{\mathcal{C}_{#1}^{'\mathrm{(eff)}}}\xspace}       
\def\Ope#1    {\ensuremath{\mathcal{O}_{#1}}\xspace}                       
\def\Opep#1   {\ensuremath{\mathcal{O}_{#1}^{'}}\xspace}                    

\newcommand{\tev}{\ifthenelse{\boolean{inbibliography}}{\ensuremath{~T\kern -0.05em eV}\xspace}{\ensuremath{\mathrm{\,Te\kern -0.1em V}}}\xspace}
\newcommand{\gev}{\ensuremath{\mathrm{\,Ge\kern -0.1em V}}\xspace}
\newcommand{\mev}{\ensuremath{\mathrm{\,Me\kern -0.1em V}}\xspace}
\newcommand{\kev}{\ensuremath{\mathrm{\,ke\kern -0.1em V}}\xspace}
\newcommand{\ev}{\ensuremath{\mathrm{\,e\kern -0.1em V}}\xspace}
\newcommand{\gevc}{\ensuremath{{\mathrm{\,Ge\kern -0.1em V\!/}c}}\xspace}
\newcommand{\mevc}{\ensuremath{{\mathrm{\,Me\kern -0.1em V\!/}c}}\xspace}
\newcommand{\gevcc}{\ensuremath{{\mathrm{\,Ge\kern -0.1em V\!/}c^2}}\xspace}
\newcommand{\gevgevcccc}{\ensuremath{{\mathrm{\,Ge\kern -0.1em V^2\!/}c^4}}\xspace}
\newcommand{\mevcc}{\ensuremath{{\mathrm{\,Me\kern -0.1em V\!/}c^2}}\xspace}

\def\pb {\ensuremath{\rm \,pb}\xspace}

\def\invfb   {\ensuremath{\mbox{\,fb}^{-1}}\xspace}

\def\gsim{{~\raise.15em\hbox{$>$}\kern-.85em
          \lower.35em\hbox{$\sim$}~}\xspace}
\def\lsim{{~\raise.15em\hbox{$<$}\kern-.85em
          \lower.35em\hbox{$\sim$}~}\xspace}

\def\sqs   {\ensuremath{\protect\sqrt{s}}\xspace}

\def\pt         {\mbox{$p_{\rm T}$}\xspace}

\newcommand{\lum} {\ensuremath{\mathcal{L}}\xspace}
\newcommand{\intlum}[1]{\ensuremath{\int\lum=#1}\xspace}  

\def\pythia     {\mbox{\textsc{Pythia}}\xspace}

\def\tell1  {TELL1\xspace}
\def\ukl1   {UKL1\xspace}

\usepackage{cite} 
\usepackage{mciteplus}

\usepackage{longtable} 
\usepackage{booktabs}

\begin{document}

\renewcommand{\thefootnote}{\fnsymbol{footnote}}
\setcounter{footnote}{1}


\begin{titlepage}
\pagenumbering{roman}

{\normalfont\bfseries\boldmath\huge
\begin{center}
Probing the  diffractive production of gauge bosons at forward rapidities
\end{center}
}

\vspace*{2.0cm}

\begin{center}
{Eduardo Basso$^1$, Victor P. Gon\c{c}alves$^{2}$, Murilo S. Rangel$^1$}\\
\vspace{2cm}
$^1$ Instituto de F\'{\i}sica, Universidade Federal do Rio de Janeiro, 
Caixa Postal 68528, Rio de Janeiro, RJ 21941-972, Brazil.\\
$^2$  Instituto de F\'{\i}sica e Matem\'atica, Universidade Federal de Pelotas, 
CEP 96010-900, Pelotas, RS, Brazil.
\end{center}

\vspace{\fill}

\begin{abstract}
  \noindent

The gauge boson production at forward rapidities in single diffractive events 
at the LHC is investigated considering $pp$ collisions at $\sqrt{s} =$ 8 and 
13 TeV. The impact of gap survival effects is analysed using two different 
models for the soft rescattering contributions. We demonstrate that using the 
Forward Shower Counter Project at LHCb -- HERSCHEL, together with the 
Vertex Locator -- VELO, it is possible to discriminate diffractive 
production of the gauge bosons \PW and \PZ from the non-diffractive processes 
and studies of the Pomeron structure and diffraction phenomenology are feasible. 
Moreover, we show that the analysis of this process can be useful to constrain 
the modelling of the gap survival effects.

\end{abstract}

\vspace*{2.0cm}

\vspace{\fill}

\end{titlepage}





\renewcommand{\thefootnote}{\arabic{footnote}}
\setcounter{footnote}{0}



\pagestyle{plain} 
\setcounter{page}{1}
\pagenumbering{arabic}


%

\section{Introduction}
\label{sec:intro}

At present days one can say that Quantum Chromodynamics factorization 
\cite{Collins:1985ue,Collins:1988ig} has been thoroughly tested so that it can 
be taken as the most powerful tool describing high energy hadronic collisions. 
Usually, the hard perturbation contributions are well separated from the soft 
non-perturbative ones, which are encoded in the parton distributions 
functions (PDF). This idea was extended long ago to the case of diffractive 
deep inelastic scattering (DDIS), for which factorization proofs have been 
carefully proved (see \cite{Collins:1997sr} and references therein). Yet the 
Regge factorization of the diffractive dissociation into a two step process, 
as suggested long ago by Ingelman and Schlein (IS) \cite{Ingelman:1984ns} and 
not proven in pQCD, have been largely used to describe hard diffractive events 
in $ep$ collisions \cite{datadifhera}. The IS approach essentially considers 
that the diffractive processes can be described in terms of the probability of 
the proton to emit a colour single object -- the  Pomeron -- and the subsequent 
interaction of the partons inside such Pomeron with the virtual photon emitted 
by the incident electron. This introduces the Pomeron parton distribution 
functions, which can be extracted from data where a hard final state is 
produced and a leading hadron is detected.

When it comes to diffractive events in $pp$ collisions, however, one have to 
be careful using these ideas, since factorization has shown to be broken when 
going from DDIS at HERA to hadron-hadron collisions at the Tevatron and the 
LHC. Indeed, theoretical studies \cite{PhysRevD.47.101,Collins:1997sr} performed before 
the experimental confirmation predicted that the breakdown of the factorization 
due to soft rescattering 
corrections associated to reinteractions (often referred to as multiple 
scatterings) between spectator partons of the colliding hadrons. These 
processes can produce additional final - state particles which fill the would-be 
rapidity gap and suppress the diffractive events. Consequently, in order to 
estimate the diffractive cross sections in hadronic collisions we need to take 
into account for the probability that such emission does not occur. One 
possible approach to treat this problem is to assume that the hard process 
occurs on a short enough timescale such that the physics that generate the 
additional particles can be factorized and accounted by an overall factor, 
denoted gap survival factor $\avg{S^2}$, multiplying the cross section 
calculated using the collinear factorization and the diffractive parton 
distributions extracted from HERA data. The modelling and magnitude of this 
factor still is a theme of intense debate \cite{kmr,Gotsman:2014pwa,roman}. In 
general the values of $\avg{S^2}$  depend on the energy, being typically of 
order $1-5$\% for LHC energies \cite{Khoze:2013dha,Gotsman:2011xc}. Such 
approach have been largely used in the literature to estimate the hard 
diffractive processes at the LHC (See e.g. Refs. \cite{magno,marta1,cristiano,
royon,royon2,marta2,cristiano2,nos_prd,marquet,marta3}) with reasonable success 
to describe the current data. On the other hand, a distinct approach to treat 
the soft rescattering interactions that destroy the rapidity gap have been 
proposed very recently \cite{Rasmussen:2015qgr}. The basic idea is to explore 
the fact that the diffractive factorization breaking effects are intimately 
related to multiple scattering in hadronic collisions. They start from the IS 
approach and adds a dynamically calculated rapidity gap survival factor, derived 
from the modelling of multiparton interactions as implemented in the \pythia 8~
\cite{Sjostrand:2014zea}. The dynamical gap survival (DGS) describes the 
factorization breaking as a function of the hard process studies and its 
kinematics. As demonstrated in Ref. \cite{Rasmussen:2015qgr}, its predictions 
are in reasonable agreement with current data.

In order to constrain the modelling of the gap survival effects and improve 
our limited understanding of diffraction it is fundamental to experimentally 
discriminate the diffractive production from the non-diffractive processes. 
The experimental signature for e.g. single diffractive production is either 
the presence of one rapidity gap in the detector or a proton tagged in the 
final state. Forward proton detectors at ATLAS and CMS experiments are in 
general available in special data taking periods with low integrated luminosity 
\cite{review_FPWG}. There are plans to operate these in nominal data taking 
with very reduced acceptance for masses smaller than few hundred GeV. On the 
other hand,  due to the lower instantaneous luminosity present at the LHCb 
experiment, it is possible to benefit  from its low pile-up conditions and its 
ability to reject activity in the backward region, to perform the study of the 
single diffractive processes at the LHCb. Such advantages have been 
demonstrated in the study of the heavy quark production in photon and Pomeron 
induced interactions performed in Ref. \cite{nos_prd}.

The LHCb detector is a single-arm spectrometer covering the forward region of 
$2<\Peta<5$~\cite{LHCb-DP-2014-002}. Recently, the LHCb collaboration 
published studies of the \PW and \PZ production inclusively~
\cite{LHCb-PAPER-2016-021,LHCb-PAPER-2014-033,LHCb-PAPER-2015-001,
LHCb-PAPER-2015-049} and in association with jets~\cite{LHCb-PAPER-2013-058,
LHCb-PAPER-2014-055,LHCb-PAPER-2015-021,LHCb-PAPER-2016-011}. These results 
show great ability of the LHCb experiment to make precise measurements of 
forward \PW and \PZ boson production. In addition, the LHCb experiment is also 
able to reject activity in the backward region using tracks reconstructed in 
the Vertex Locator (VELO) sub-detector. The experiment runs at lower 
instantaneous luminosity benefiting from low pile-up conditions. Indeed, LHCb 
has already published central exclusive analyses exploring the ability of 
requiring a backward gap~\cite{lhcbYcep,lhcbJPSIcep,lhcbDDcep} in RunI. New 
results of LHCb central exclusive production has also been available using the 
new forward shower counters \cite{LHCb-DP-2016-003} that extends the pseudorapidity region in which 
charged particles can be vetoed~\cite{Jpsi13tev}. Additionally, at the Run II, 
it is possible to study diffractive events at the LHCb experiment by selecting 
the muon within the LHCb acceptance and require no particles in the backward 
region of $-3.5<\eta<-1.5$ (VELO) and $-8.0<\eta<-5.5$ (HERSCHEL).

In this paper we study the gauge boson production in single diffractive events 
and propose to use the HERSCHEL, together with the VELO, to discriminate diffractive 
production of the gauge bosons \PW and \PZ from the non-diffractive processes. 
The results presented here indicate that studies of the Pomeron structure and 
diffraction phenomenology are feasible at the LHC and that the analysis of 
this process can be useful to constrain the modelling of the gap survival 
effects. The paper is organized as follows. In the next Section we present a 
brief review of the formalism used for the gauge boson production in single 
diffractive events. In Section~\ref{sec:results} we present our results for the 
total cross sections and pseudorapidity distributions. We consider the \PW and 
\PZ production as well as its production associated to a jet and estimate the 
ratio between cross sections, which cancel several of the experimental and 
theoretical systematic uncertainties, considering $pp$ collisions at 
$\sqrt{s} =$ 8 and 13 TeV and two distinct models for the gap survival effects. 
Finally, in Section~\ref{sec:discussion} we summarize our main conclusions.

\section{Single diffractive gauge boson production}
\label{sec:Theory}

\begin{figure}[tbp]
\centering 
\includegraphics[width=.45\textwidth]{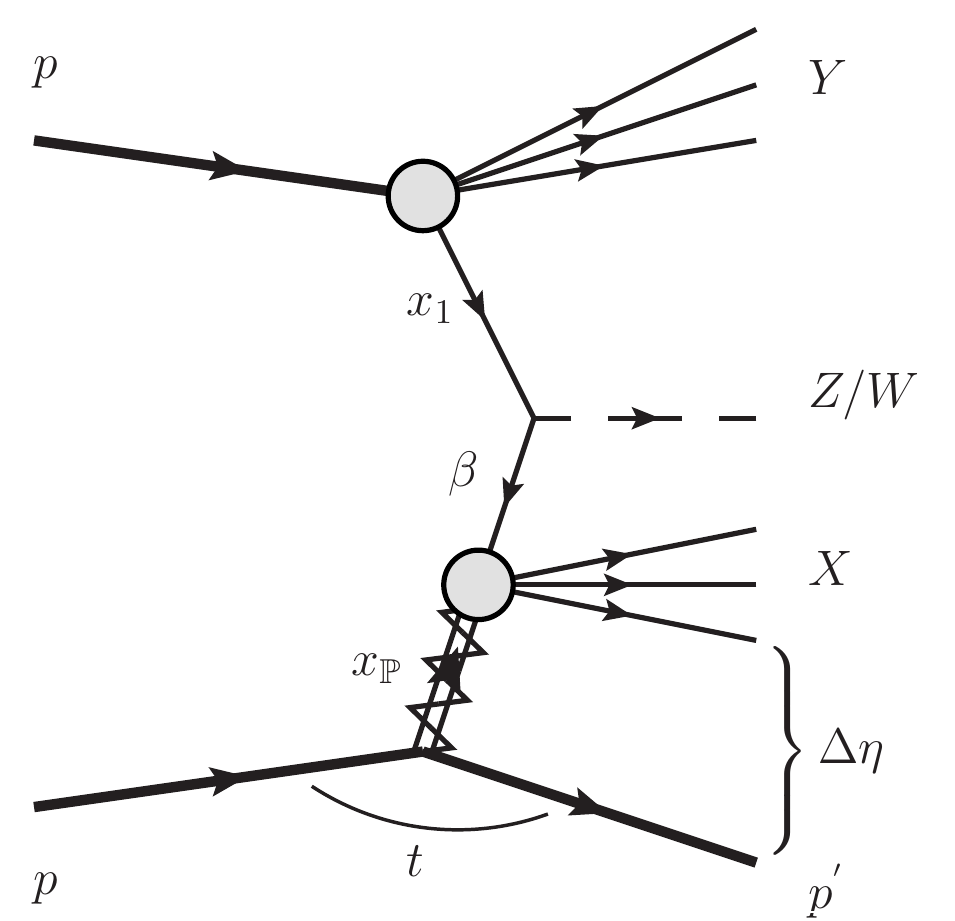}
\caption{\label{fig:wzsd} LO diagram contributing to single diffractive 
production of gauge bosons.}
\end{figure}

In the following we apply the IS approach \cite{Ingelman:1984ns} for the 
diffractive gauge boson production represented in Fig.~\ref{fig:wzsd}. This 
model, denoted often Resolved Pomeron Model, assumes that the Pomeron has a 
well defined partonic structure and that the hard process takes place in a 
Pomeron - proton or proton - Pomeron interaction in the case of single 
diffractive processes. At leading order the gauge boson production  is 
determined by the annihilation processes $q \bar q \rightarrow G$ ($G = W$ 
or $Z$). Higher order contributions are not considered here and can be taken 
into account effectively by a $K$ factor. In order to estimate the hadronic 
cross sections we have to convolute the cross section for this partonic 
subprocess with the inclusive and diffractive parton distribution functions. 
In the collinear factorization formalism, the single diffractive gauge boson 
production cross section can be expressed by
\begin{eqnarray}
\sigma^{SD}(pp \rightarrow G p) = \sum_{a,b} \int dx_a \int dx_b 
\left[f^D_a(x_a,\mu^2) f_b(x_b,\mu^2) + f_a(x_a,\mu^2)f^D_b(x_b,\mu^2)\right] 
\, \hat{\sigma}_{ab \rightarrow G} \,\,, \label{cross} 
\end{eqnarray}
where $f_i(x_i,\mu^2)$ and $f_i^D(x_i,\mu^2)$ are the inclusive and diffractive 
parton distribution functions, respectively, and we included both $p \Pom$ and 
$\Pom p$ interactions. Moreover, $\hat{\sigma}_{ab\rightarrow G}$ denotes the 
hard partonic interaction producing a gauge boson.

The Resolved Pomeron Model considers the difractive parton distributions as a 
convolution of the Pomeron flux emitted by the proton, $f_{\Pom}(x_{\Pom})$, 
and the parton distributions in the Pomeron, $g_{\Pom}(\beta, \mu^2)$, 
$q_{\Pom}(\beta, \mu^2)$, where $\beta$ is the momentum fraction carried by the 
partons inside the Pomeron. The Pomeron flux is given by 
$f_{\Pom}(x_{\Pom})= \int_{t_{min}}^{t_{max}} dt f_{\Pom/p}(x_{{\Pom}}, t)$, 
where 
\begin{equation}
\label{eq:h1flux}
f_{\Pom/p}(x_\Pom,t) = A_\Pom\frac{e^{B_\Pom t}}{x_\Pom^{2\alpha_\Pom(t) - 1}}\,.
\end{equation}
The normalization of the flux is such that the relation 
$x_\Pom \int^{t_{\text{min}}}_{{t_\text{cut}}} \d t f_{\Pom/p}(x_\Pom,t) = 1$ 
holds at $x_\Pom = 0.003$, where $|t_\text{cut}| = 1$ GeV is limited by the 
measurement and $|t_\text{min}| \simeq (m_\text{p} x_\Pom)^2/(1-x_\Pom)$ is the 
kinematic limit of accessible momentum $|t|$. The Pomeron flux factor is 
motivated by Regge theory, where the Pomeron trajectory assumed to be linear, 
$\alpha_{\Pom} (t)= \alpha_{\Pom} (0) + \alpha_{\Pom}^\prime t$, and the 
parameters $B_{\Pom}$, $\alpha_{\Pom}^\prime$ and their uncertainties are 
obtained from fits to H1 data  \cite{Aktas:2006hy, Aktas:2006hx}. 
The diffractive quark and gluon distributions are then given by
\begin{eqnarray}
{ q^D(x,\mu^2)}=\int dx_{\Pom}d\beta \delta (x-x_{\Pom}\beta)f_{\Pom}(x_{\Pom})
q_{\Pom}(\beta, \mu^2)={ \int_x^1 \frac{dx_{\Pom}}{x_{\Pom}} f_{\Pom}(x_{\Pom}) 
q_{\Pom}\left(\frac{x}{x_{\Pom}}, \mu^2\right)}\,, \\
{ g^D(x,\mu^2)}=\int dx_{\Pom}d\beta \delta (x-x_{\Pom}\beta)f_{\Pom}(x_{\Pom})
g_{\Pom}(\beta, \mu^2)={ \int_x^1 \frac{dx_{\Pom}}{x_{\Pom}} f_{\Pom}(x_{\Pom}) 
g_{\Pom}\left(\frac{x}{x_{\Pom}}, \mu^2\right)}\,.
\end{eqnarray}
In the present analysis the H1 Fit A is used \cite{Aktas:2006hy, Aktas:2006hx}, 
which has the slope parameter set to $B_\Pom = 5.5$ GeV${}^{-2}$ and 
$\alpha_\Pom^\prime = 0.06$ GeV${}^{-2}$, while $\alpha_\Pom(0) = 1.118 \pm 
0.008 $. Moreover, we use the inclusive parton distributions as given by the 
CT10 parametrization \cite{ct10}. It is important to emphasize that at large 
values of the Pomeron longitudinal $x_{\Pom}$, subleading contributions 
associated to Reggeon exchange can be important in some regions of the phase 
space (For a recent discussion see e.g. Ref. \cite{marquet}). In what follows 
we disregard this contribution and postpone for a future study the analysis 
of its impact.

Recently, the Resolved Pomeron Model described above have been implemented in 
\pythia 8 \cite{Rasmussen:2015qgr}, allowing to estimate the hard diffractive 
events at the LHC using this event generator. However, in order to describe 
the data for diffractive events in $pp$ collisions we should taken into account 
the soft rescattering contributions discussed in the Introduction and that 
imply the breakdown of the diffractive factorization. In the last years, many 
models have been proposed to describe the Gap Survival Probability (GSP). A 
simplistic approach is based on the assumption that the soft rescattering 
contributions can be factorized from the hard processes and can be taken into 
account as a multiplicative factor to the cross section. Such models were 
shortly summarized recently in \cite{Gotsman:2014pwa}, with its value being 
typically of order $1-5$\% for LHC energies, according to the calculations 
of the Durham (2 channel eikonal model) \cite{Khoze:2013dha} and Tel-Aviv 
\cite{Gotsman:2011xc} groups. On the other hand, there are a few approaches 
aiming to model the GSP as a dynamical process \cite{Edin:1995gi,
Buchmuller:1995qa,Brodsky:2002ue,Brodsky:2004hi,Pasechnik:2010zs,
Ingelman:2015qrt}. In particular, in Ref. \cite{Rasmussen:2015qgr} the authors 
have proposed to use the full machinery of Multiple Parton Interactions (MPI) 
present in the \pythia 8 generator to account for the absorptive corrections 
upon which the GSP are built. The basic idea is that the MPI that occur 
between the incoming hadrons can create colour flows that fill with hadrons 
the rapidity range available, destroying the rapidity gap associated to single 
diffractive events. 
Since MPI are dynamically generated, the GSP obtained from such method is also 
dynamical. When this framework is switched off, ``pure'' diffractive events are 
generated and the factorization breaking events should be additionally included 
through a multiplicative factor. In what follows we will consider both 
prescriptions to treat the GSP, denoting by SD1 those derived assuming that 
$\avg{S^2} = 0.05$ at the LHC energies, as in Refs. \cite{magno,marta1,
cristiano,royon,royon2,marta2,cristiano2,nos_prd,marquet,marta3}, and by SD2 
those associated the dynamically generated gap survival based on MPI as 
implemented in \pythia 8 \cite{Rasmussen:2015qgr}. The comparison between these 
two models for the GSP allows to estimate the current theoretical 
uncertainty in the predictions for diffractive events in comparison to the 
non-diffractive (ND) events, which will be studied here as well.

\section{Results}
\label{sec:results}

In what follows we present results for the \PW and \PZ production in $pp$ 
collisions at $\sqs=8$ and $13 \tev$. The main focus will be in forward region 
covered by the LHCb detector. The events have been generated in \pythia 8 and 
we have selected  the muon within the LHCb acceptance and required no particles 
in the backward region of $-3.5<\eta<-1.5$ (VELO) and $-8.0<\eta<-5.5$ 
(HERSCHEL). The VELO gap requirement is performed using charged particles with 
momentum greater than $100\mev$. We assumed that the  HERSCHEL is able to detect 
charged and neutral with $\pt>0.5\gev$. Additionally, the gauge boson selection 
used in Ref.~\cite{LHCb-PAPER-2016-011} have been applied, i.e., the \PW boson 
is identified using its decay to a muon and a neutrino and the Z boson is 
identified using its decay to a muon pair. The muon must have $\pt(\Pmu)>20\gev$ 
and $2.0<\eta(\Pmu)<4.5$. 

\begin{table}[t]
\begin{center}
\bgroup
\def\arraystretch{1.3}
\setlength\tabcolsep{1.5pt}
\begin{tabular}{l|l|l||l|l}
\toprule
 ~ &  \multicolumn{2}{c|}{$\sqs=13\tev$} &  \multicolumn{2}{c}{$\sqs=8\tev$} \\
\midrule
Process ~ & ~ No gap ~ & ~ VH gap ~ & ~ No gap & ~ V gap \\
\midrule
\Wp ND   ~ & ~ $1.7\times10^3$ ~ & ~ $4.4 $ ~ & ~  $1.4\times10^3$ ~& ~$15  $   \\
\Wp SD1  ~ & ~ $15           $ ~ & ~ $1.7 $ ~ & ~  $12           $ ~& ~$1.6 $   \\
\Wp SD2  ~ & ~ $48           $ ~ & ~ $5.6 $ ~ & ~  $41           $ ~& ~$5.1 $   \\
\midrule
\Wm ND   ~ & ~ $1.2\times10^3$ ~ & ~ $3.6 $ ~ & ~  $1.1\times10^3$ ~& ~$12  $   \\
\Wm SD1  ~ & ~ $10           $ ~ & ~ $0.9 $ ~ & ~  $9.4          $ ~& ~$0.94$   \\
\Wm SD2  ~ & ~ $35           $ ~ & ~ $3.0 $ ~ & ~  $32           $ ~& ~$3.1 $   \\
\midrule
\Z  ND   ~ & ~ $150          $ ~ & ~ $0.38$ ~ & ~  $100          $ ~& ~$1.3 $   \\
\Z  SD1  ~ & ~ $1.4          $ ~ & ~ $0.15$ ~ & ~  $0.9          $ ~& ~$0.12$   \\
\Z  SD2  ~ & ~ $4.4          $ ~ & ~ $0.50$ ~ & ~  $3.0          $ ~& ~$0.42$   \\
\bottomrule
\end{tabular}
\egroup
\caption{\label{tab:cs} Cross-sections in \pb for \PW and \PZ production in 
the LHCb fiducial region before and after requiring a region void of particles 
for 8 and 13 TeV. For 13 TeV the gap requirement considers the VELO and 
HERSCHEL detectors (VH gap) while for 8 TeV only VELO detector is used (V gap).}
\end{center}
\end{table}

\begin{figure}[tbp]
\centering 
\includegraphics[width=.45\textwidth]{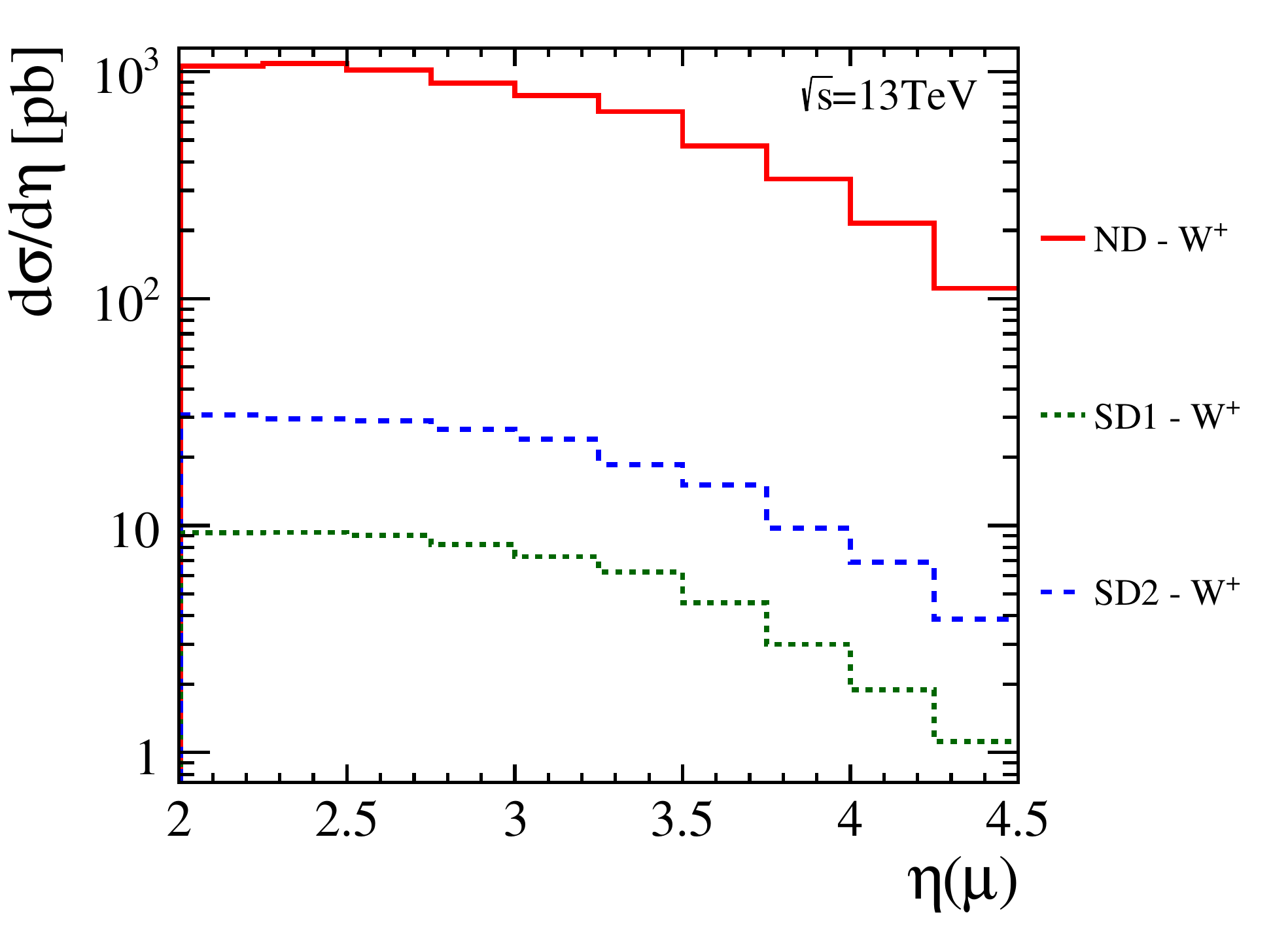}
\includegraphics[width=.45\textwidth]{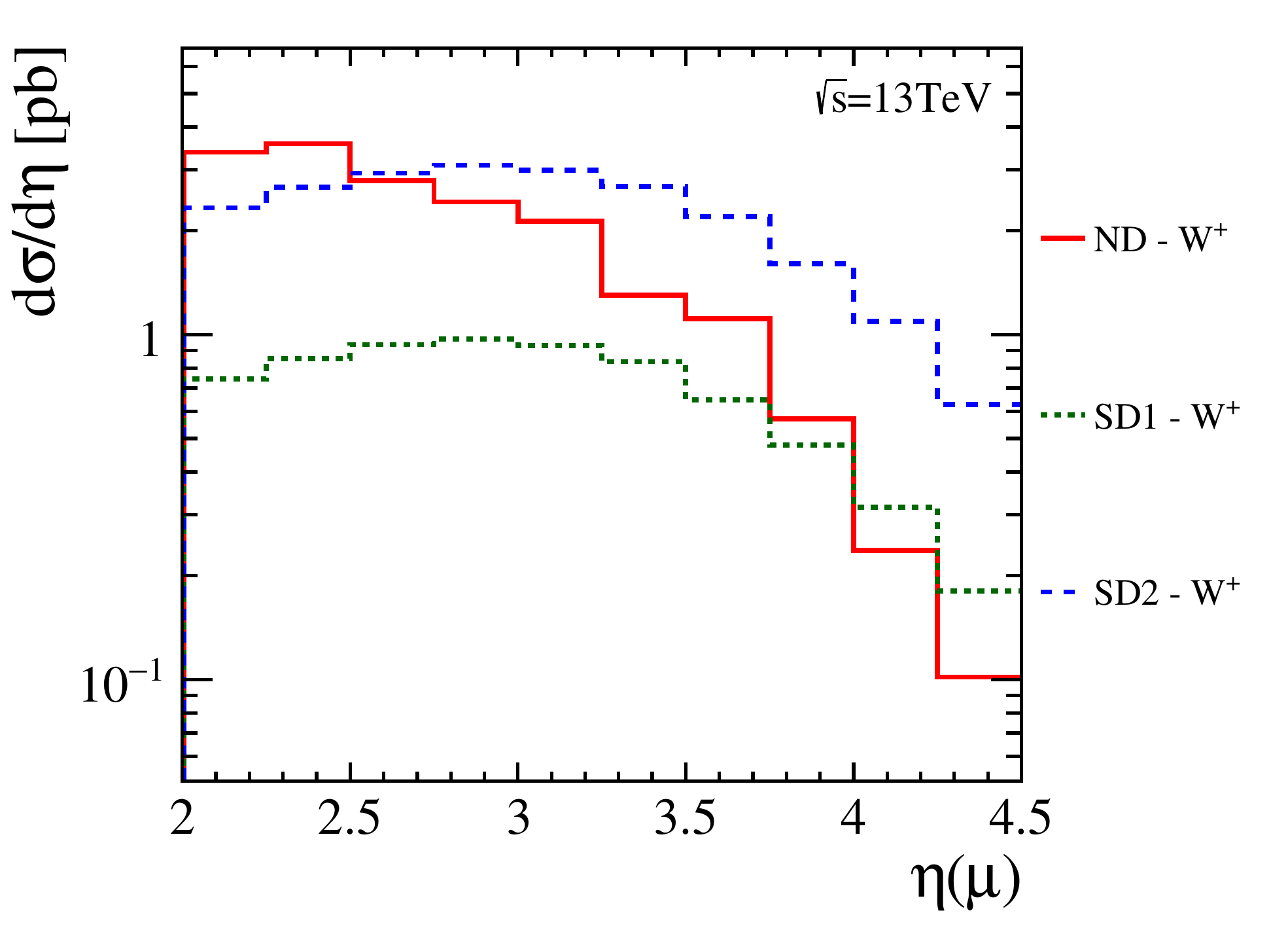}
\hfill
\caption{\label{fig:dsigdeta} Differential cross-section as function of 
$\eta(\Pmu)$ for non-diffractive and single-diffrative production of \PW in 
$pp$ collisions at $\sqrt{s}=$ 13 TeV before (left) and after (right) the gap 
requirement is assumed.}
\end{figure}

Predictions for the cross-sections in the LHCb fiducial region are 
presented in Table~\ref{tab:cs} for $pp$ collisions at 8 and $13\tev$. 
We present the results obtained before and after requiring a region void of 
particles (gap) in the final state. For 13 TeV the gap requirement considers 
the VELO and HERSCHEL detectors (denoted VH gap) while for 8 TeV only VELO 
detector is used (denoted V gap). As expected, we have that if the gap 
requirement is not assumed, the non-diffractive (ND) production of gauge 
bosons is dominant, being two orders of magnitudes larger than the single 
diffraction (SD) one.  On the other hand, if a gap is required, the ND 
contribution is strongly suppressed and becomes of the order of the SD process. 
In particular, such suppression is larger at $\sqrt{s} = 13$ TeV, which 
demonstrates the impact of the HERSCHEL detector for the study of diffractive 
events. Regarding the models for the gap survival, we have that its predictions 
are similar, with the SD1 model predicting cross sections that are a factor 3 
smaller than the SD2 one. An important aspect is that the SD2 model implies 
that the single diffraction production of $W^+$ and $Z$ becomes dominant at 
13 TeV. This results motivates the analysis of differential distributions that 
can be measured experimentally. In Fig.~\ref{fig:dsigdeta} we present the 
predictions for the differential cross-section as function of the pseudorapidity 
of the muon $\eta(\Pmu)$  for the \Wp production. In the left (right) panel we 
show the results obtained before (after) the gap requirement is applied. One can 
see that the additional HERSCHEL gap requirement improves the discrimination 
between non-diffractive and diffrative processes for $13\tev$ prediction and 
implies that SD contribution becomes dominant in a large range of 
pseudorapidities in the case of the SD2 model.

\begin{figure}[tbp]
\centering 
\includegraphics[width=.45\textwidth]{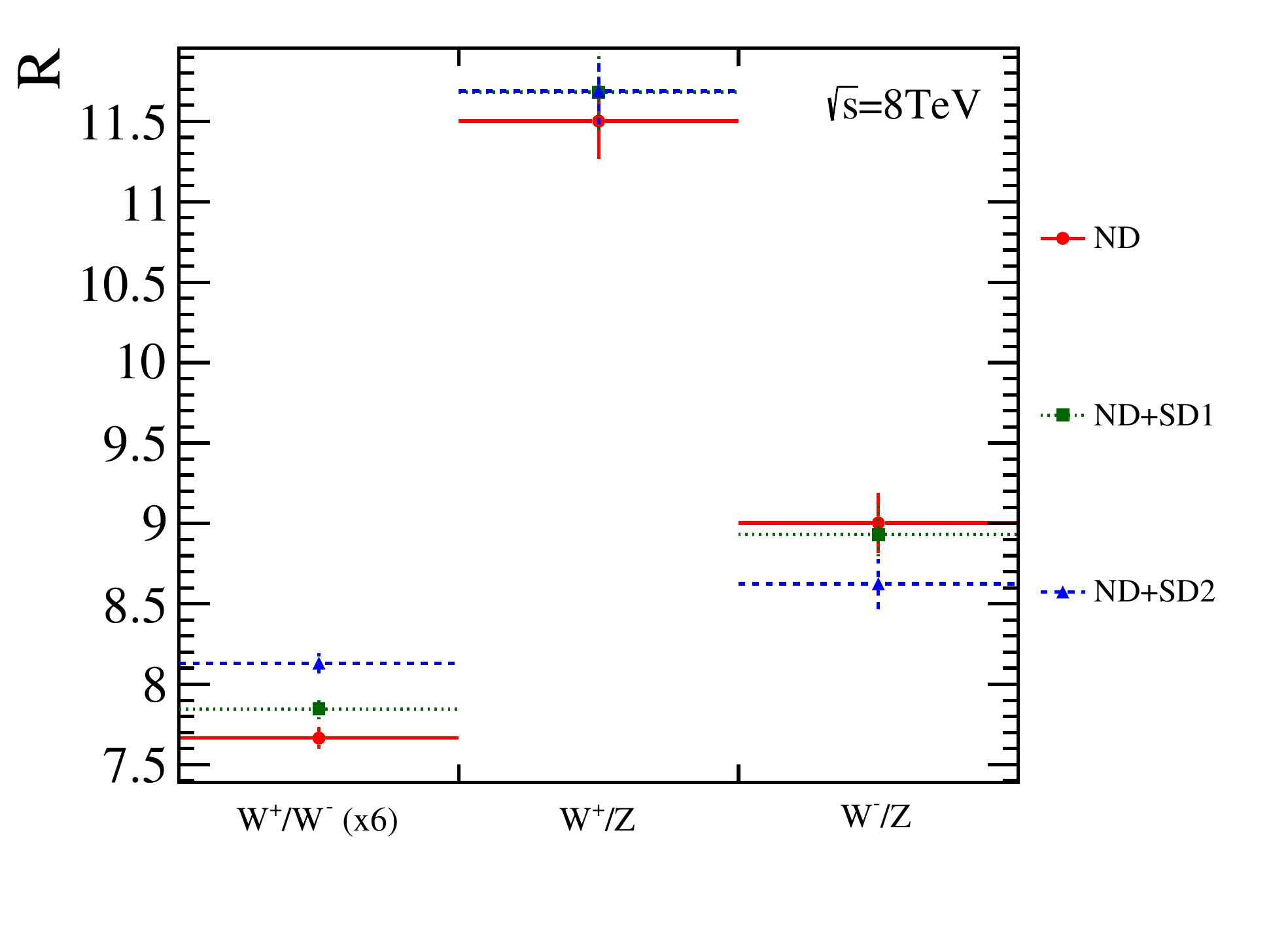}
\includegraphics[width=.45\textwidth]{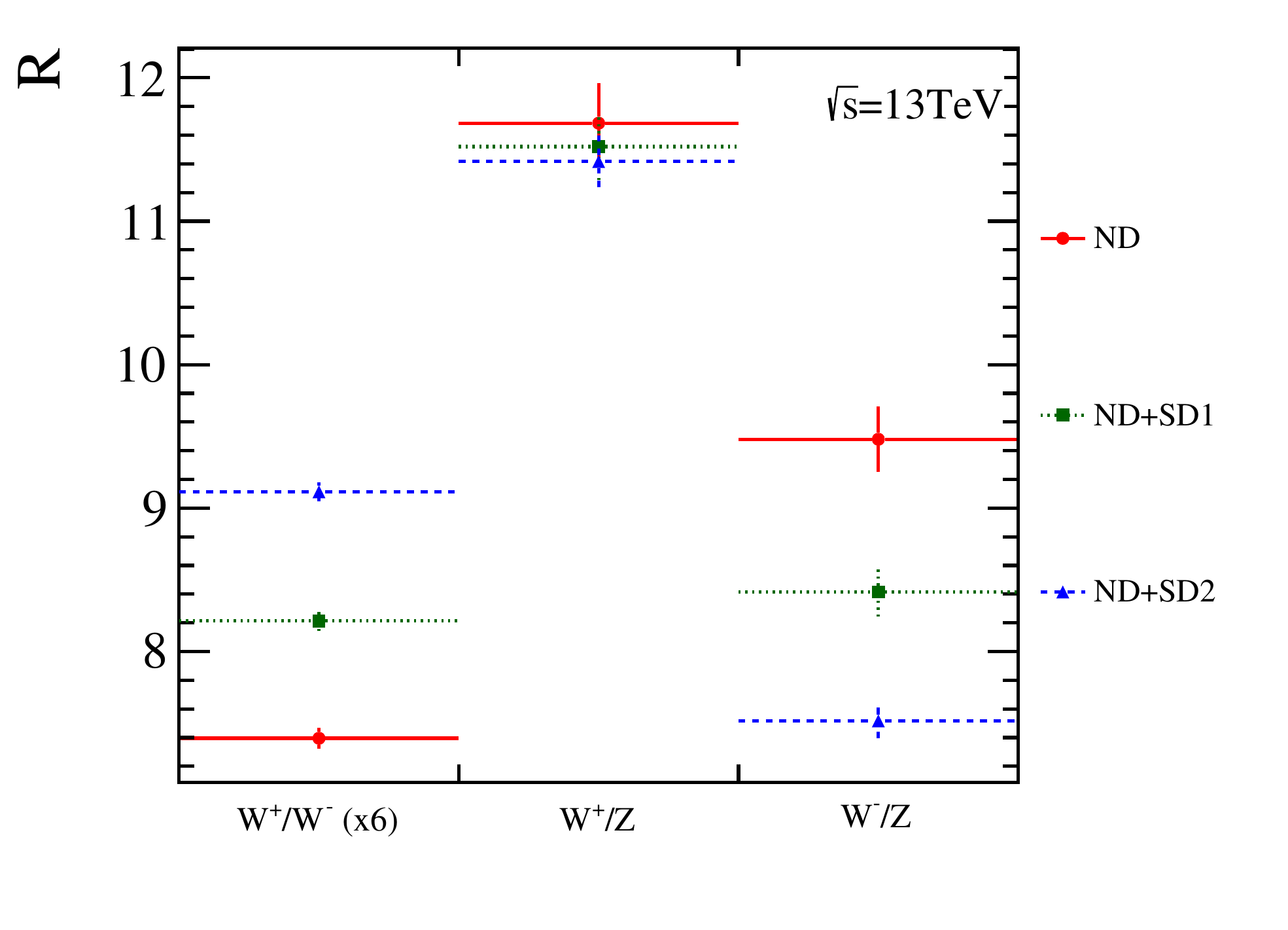}
\caption{\label{fig:ratios} Ratio of cross-sections for \PW and \PZ production: 
$\rwpm=\frac{\sigma_\Wp}{\sigma_\Wm}$, $\rwmz=\frac{\sigma_\Wm}{\sigma_\PZ}$ 
and $\rwpz=\frac{\sigma_\Wp}{\sigma_\PZ}$. The displayed errors are the 
expected statistical errors considering \intlum 5 \invfb for 13 TeV and \intlum 
2 \invfb for 8 TeV. $\rwpm$ is multiplied by 6.}
\end{figure}

In order to minimize possible experimental and theoretical systematic 
uncertainties we propose the analysis of the ratios between the cross sections 
defined by $\rwpm=\frac{\sigma_\Wp}{\sigma_\Wm}$, 
$\rwmz=\frac{\sigma_\Wm}{\sigma_\PZ}$ and $\rwpz=\frac{\sigma_\Wp}{\sigma_\PZ}$. 
The error in the ratios \rwmz and \rwpz is dominated by sample size. As we are 
interested in investigate the impact of the single diffraction contribution we 
will estimate these ratio considering only the ND contribution and will 
compare these with those obtained summing the ND and SD contributions. 
Assuming an integrated luminosity of \intlum 5 \invfb for 13 TeV and \intlum 2 
\invfb for 8 TeV, we have estimated these ratios and the predictions with 
their respective expected statistical errors are shown in Fig.~\ref{fig:ratios}. 
One have that for $\sqrt{s} =$ 8 TeV the impact of the SD contribution is 
smaller than 10 \%. On the other hand,  for $\sqrt{s} =$ 13 TeV and for the 
ratios $\rwpm=\frac{\sigma_\Wp}{\sigma_\Wm}$ and 
$\rwpz=\frac{\sigma_\Wp}{\sigma_\PZ}$, one have that its contribution can be 
of the order of 20 \% and is sensitive to the model used for the description 
of the gap survival effects.

\begin{table}[t]
\begin{center}
\bgroup
\def\arraystretch{1.3}
\setlength\tabcolsep{1.5pt}
\begin{tabular}{l|l|l||l|l}
\toprule
 ~ &  \multicolumn{2}{c|}{$\sqs=13\tev$} &  \multicolumn{2}{c}{$\sqs=8\tev$} \\
\midrule
Process ~ & ~ No gap ~ & ~ VH gap ~ & ~ No gap & ~ V gap \\
\midrule
\Wp$j$ ND   ~ & ~ $3.3\times10^2$ ~ & ~ $0.37$ ~ & ~  $2.1\times10^2$ ~& ~$1.2   $   \\
\Wp$j$ SD1  ~ & ~ $2.5          $ ~ & ~ $0.2 $ ~ & ~  $1.6          $ ~& ~$0.15  $   \\
\Wp$j$ SD2  ~ & ~ $8.2          $ ~ & ~ $0.65$ ~ & ~  $5.5          $ ~& ~$0.49  $   \\
\midrule
\Wm$j$ ND   ~ & ~ $2.3\times10^2$ ~ & ~ $0.29$ ~ & ~  $1.5\times10^2$ ~& ~$1.0   $   \\
\Wm$j$ SD1  ~ & ~ $1.6          $ ~ & ~ $0.1 $ ~ & ~  $1.1          $ ~& ~$0.08  $   \\
\Wm$j$ SD2  ~ & ~ $5.4          $ ~ & ~ $0.35$ ~ & ~  $3.7          $ ~& ~$0.29  $   \\
\midrule
\Z$j$ ND   ~ & ~ $31           $ ~ & ~ $0.03$ ~ & ~  $17           $ ~& ~$0.12  $   \\
\Z$j$ SD1  ~ & ~ $0.25         $ ~ & ~ $0.02$ ~ & ~  $0.14         $ ~& ~$0.01  $   \\
\Z$j$ SD2  ~ & ~ $0.86         $ ~ & ~ $0.07$ ~ & ~  $0.45         $ ~& ~$0.05  $   \\
\bottomrule
\end{tabular}
\egroup
\caption{\label{tab:csjet} Cross-sections in \pb for \PW/\PZ plus jets 
production before and after requiring a region void of particles for 
$\sqrt{s} = $ 8 and 13 TeV. For 13 TeV the gap requirement considers the 
VELO and HERSCHEL detectors (VH gap) while for 8 TeV only VELO detector 
is used (V gap).}
\end{center}
\end{table}

\begin{figure}[t]
\centering 
\includegraphics[width=.45\textwidth]{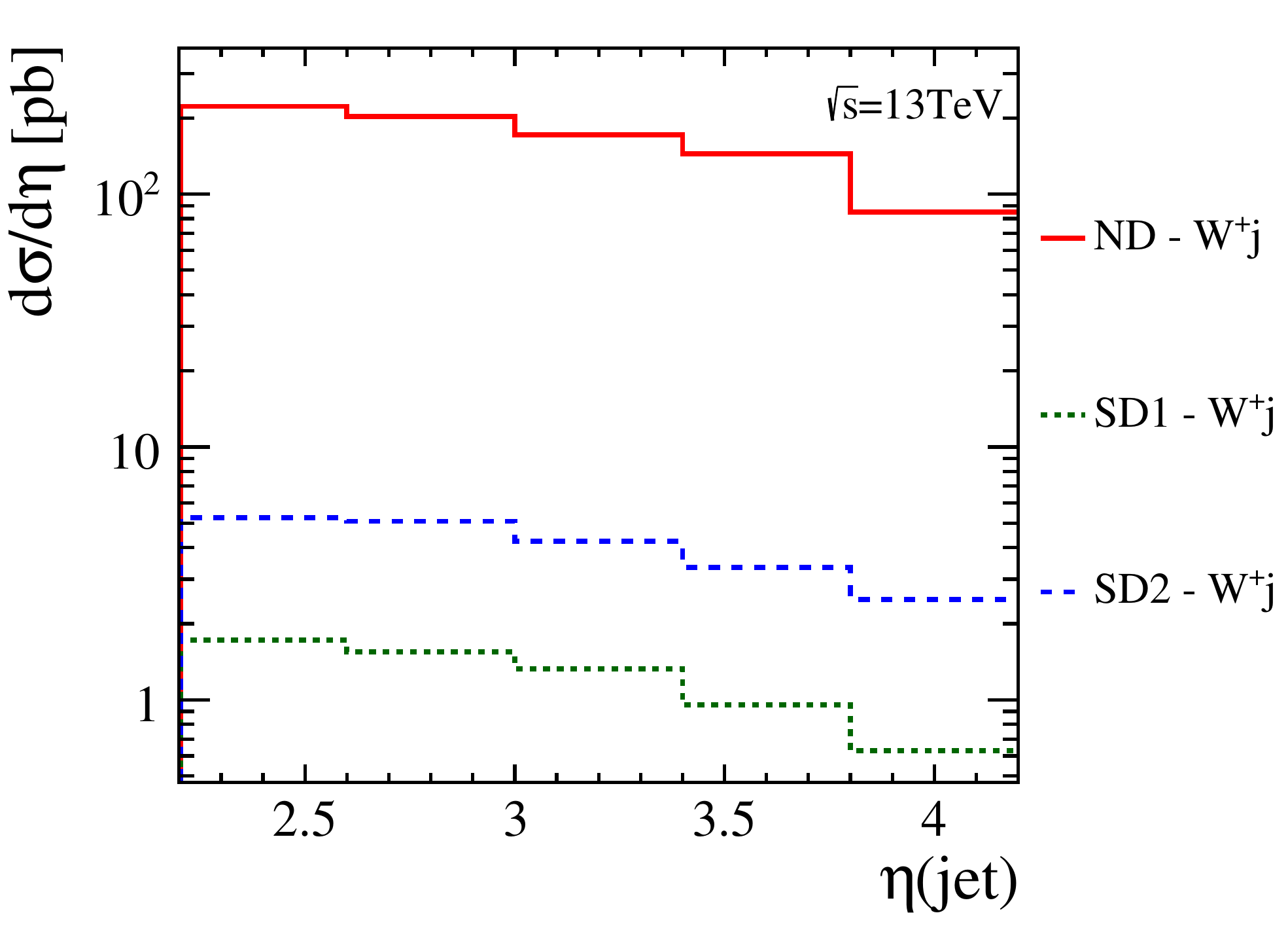}
\includegraphics[width=.45\textwidth]{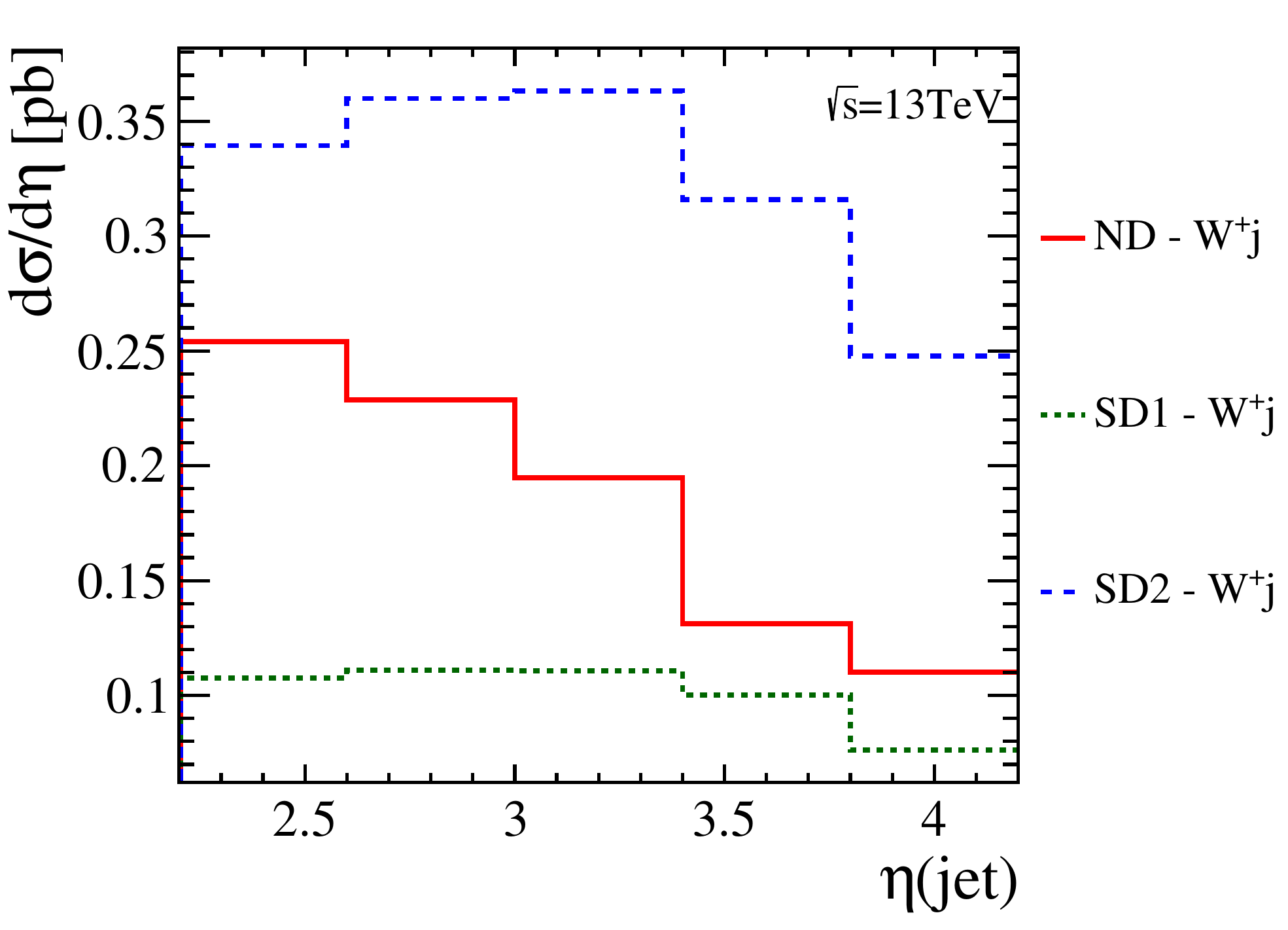}
\hfill
\caption{\label{fig:dsigdetajet} Differential cross-section as function 
$\eta({\rm jet})$ for non-diffractive and single-diffrative production of \Wp 
plus jets in $pp$ collisions at $\sqrt{s}$ = 13 TeV before (left) and after 
(right) the implementation of the gap requirement.}
\end{figure}

In addition to the study of the inclusive gauge boson production, which is 
strongly dependent on the quark distribution of the proton and of the Pomeron, 
the associate production with jets can be useful since it also is sensitive to 
the gluon distribution. In order to evaluate the impact of single diffractive 
contribution in gauge boson plus jets at LHCb, we have reconstructed the jet 
using the \antikt algorithm~\cite{antikt} with distance parameter $R=0.5$ as 
implemented in the \fastjet software package~\cite{fastjet}. In the \PW/\PZ 
plus jet selection, jets were required to have $\pt(\text{jet})>10\gev$, 
$2.2<\eta(\text{jet})<4.2$ and $\Delta\text{\it R}(\muon,\text{jet})>0.5$, 
where $\Delta\text{\it R}=\sqrt{\Delta\varphi^2+\Delta\eta^2}$. The results for 
the cross-sections in the LHCb fiducial region are presented in 
Table~\ref{tab:csjet} for 8 and $13\tev$ with and without gap requirements. 
As in the inclusive gauge boson production, the ND contribution is dominant in 
the gauge boson + jet production if a gap is not required in the final state 
and highly suppressed if the gap is required. Moreover, the SD2 model predicts 
the dominance of the single diffraction production at $\sqrt{s} = 13$ TeV if 
the gap requirement considers the VELO and HERSCHEL detectors. This behaviour 
is also present in the differential distributions. As a example, in 
Fig. ~\ref{fig:dsigdetajet} we present predictions for the \Wp plus jet 
production in $pp$ collisions at $\sqrt{s} = 13$ TeV before (left panel) and 
after (right panel) the implementation of the gap requirement. One have that 
in the case of the SD2 model, the single diffractive \Wp plus jet production 
is dominant for all values of pseudorapidity covered by the LHCb detector.

\begin{figure}[tbp]
\centering 
\includegraphics[width=.45\textwidth]{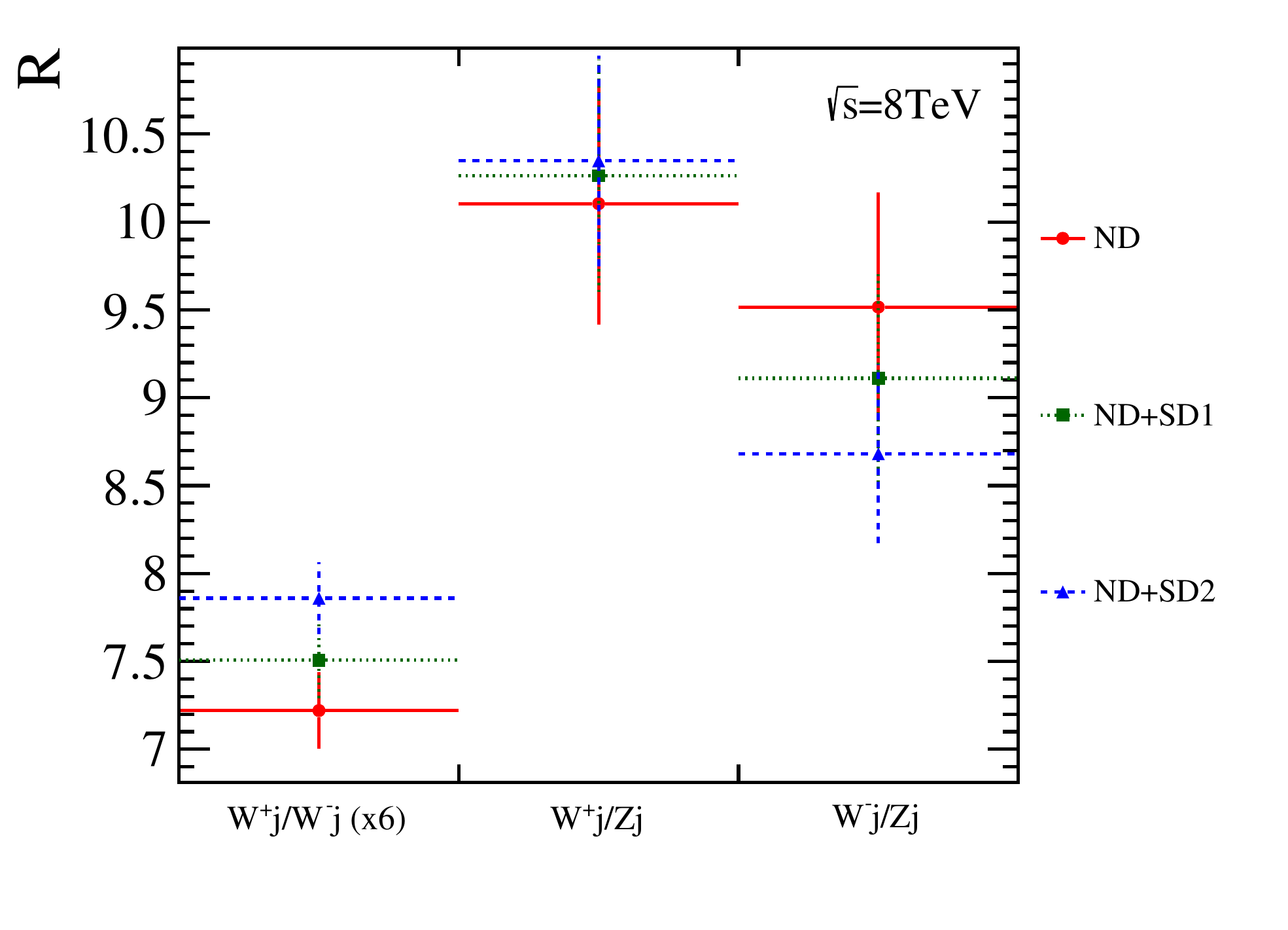}
\includegraphics[width=.45\textwidth]{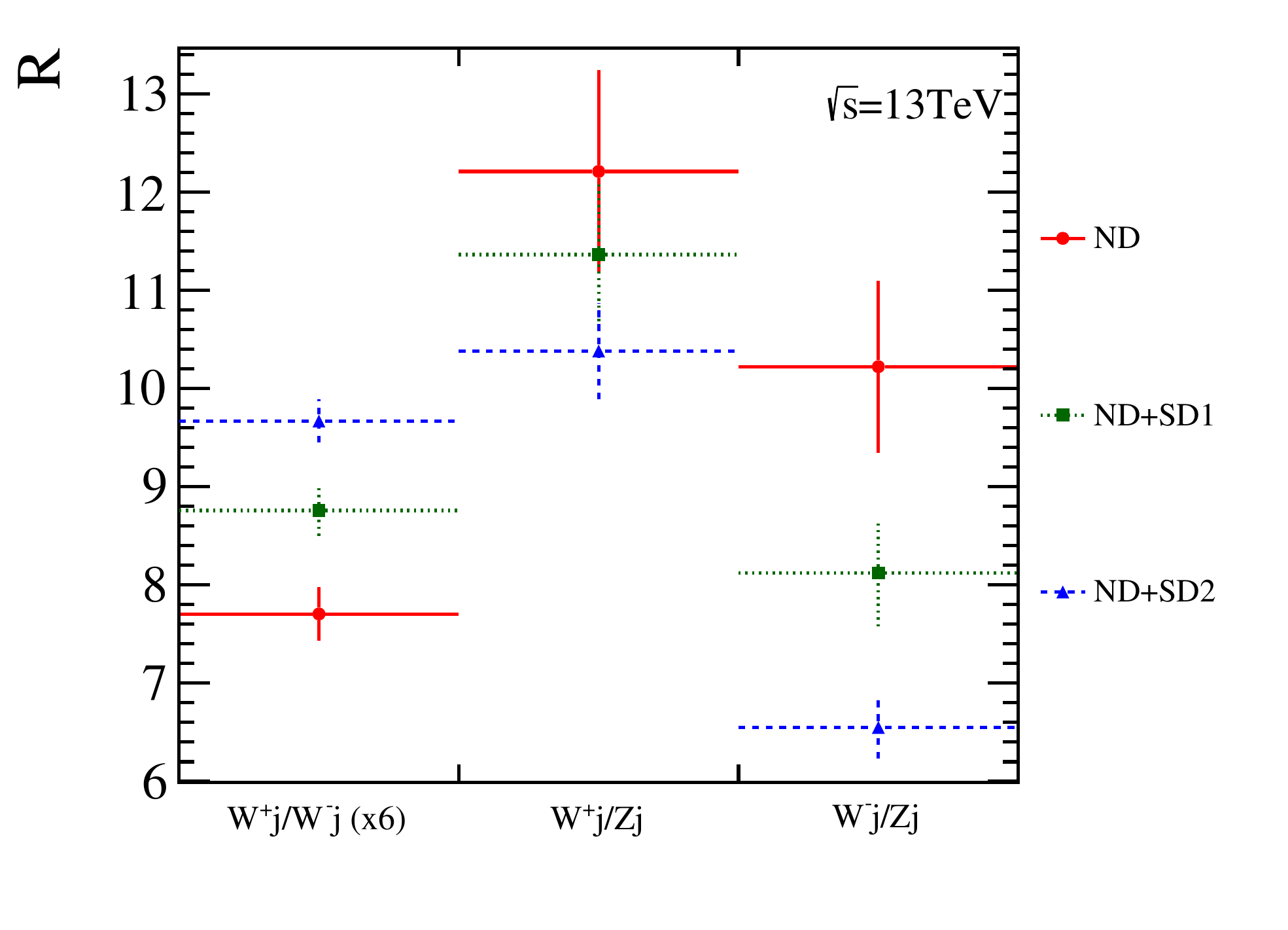}
\caption{\label{fig:ratiosjet} Ratio of cross-sections for \PW and \PZ 
production in association with jets: 
$R_1 =\frac{\sigma_{\Wp j}}{\sigma_{\Wm j}}$, 
$R_2 =\frac{\sigma_{\Wm j}}{\sigma_{\PZ j}}$ and 
$R_3=\frac{\sigma_{\Wp j}}{\sigma_{\PZ j}}$. The displayed errors are the 
expected statistical errors considering \intlum 5 \invfb for 13 TeV and \intlum 
2 \invfb for 8 TeV. $R_1$ is multiplied by 6.}
\end{figure}

Finally, in Fig.~\ref{fig:ratiosjet} we present the predictions for 
$R_1 =\frac{\sigma_{\Wp j}}{\sigma_{\Wm j}}$, 
$R_2 =\frac{\sigma_{\Wm j}}{\sigma_{\PZ j}}$ and 
$R_3=\frac{\sigma_{\Wp j}}{\sigma_{\PZ j}}$. As in the inclusive production, 
one have that these ratios are sensitive to the presence of the single 
diffractive events and its magnitude depends of the model used to describe the 
gap survival effects.

\section{Summary}
\label{sec:discussion}

In this paper we investigated the gauge boson production at forward rapidities 
in single diffractive events at the LHC. Using the \pythia 8 event generator 
we have estimated the cross sections and differential distributions for the 
\Wp, \Wm and $Z$ production, as well for the gauge boson production in 
association with jets. We have considered realistic cuts and gap requirements 
that can be performed by the LHCb Collaboration.  The present analysis is 
complementary to the studies involving the planned proton tagging detectors by 
CMS and ATLAS collaborations, with the advantage that LHCb is already setup for 
detection of diffractive events. Our results demonstrated that using the 
HERSCHEL, in addition to the VELO, it is possible to discriminate diffractive 
production of the 
gauge bosons \PW and \PZ (with and without extra jets) from the non-diffractive 
processes. As a consequence, it is possible to use the resulting experimental 
data to study in more detail the treatment of the diffractive processes. In 
particular, our results shown that the analysis of the cross sections, 
differential distributions and the ratio between cross sections can be useful 
to constrain the model used for the soft rescattering corrections that 
breakdown the diffractive factorization in hadronic collisions.

\FloatBarrier

\section*{Acknowledgements} 
  This research was supported by CNPq, CAPES, FAPERJ and FAPERGS, Brazil. 

\addcontentsline{toc}{section}{References}
\setboolean{inbibliography}{true}
\bibliographystyle{main}
\bibliography{main}

\ifx\mcitethebibliography\mciteundefinedmacro
\PackageError{main.bst}{mciteplus.sty has not been loaded}
{This bibstyle requires the use of the mciteplus package.}\fi
\providecommand{\href}[2]{#2}
\begin{mcitethebibliography}{10}
\mciteSetBstSublistMode{n}
\mciteSetBstMaxWidthForm{subitem}{\alph{mcitesubitemcount})}
\mciteSetBstSublistLabelBeginEnd{\mcitemaxwidthsubitemform\space}
{\relax}{\relax}

\bibitem{Collins:1985ue}
J.~C. Collins, D.~E. Soper, and G.~F. Sterman,
  \ifthenelse{\boolean{articletitles}}{\emph{{Factorization for Short Distance
  Hadron - Hadron Scattering}},
  }{}\href{http://dx.doi.org/10.1016/0550-3213(85)90565-6}{Nucl.\ Phys.\
  \textbf{B261} (1985) 104}\relax
\mciteBstWouldAddEndPuncttrue
\mciteSetBstMidEndSepPunct{\mcitedefaultmidpunct}
{\mcitedefaultendpunct}{\mcitedefaultseppunct}\relax
\EndOfBibitem
\bibitem{Collins:1988ig}
J.~C. Collins, D.~E. Soper, and G.~F. Sterman,
  \ifthenelse{\boolean{articletitles}}{\emph{{Soft Gluons and Factorization}},
  }{}\href{http://dx.doi.org/10.1016/0550-3213(88)90130-7}{Nucl.\ Phys.\
  \textbf{B308} (1988) 833}\relax
\mciteBstWouldAddEndPuncttrue
\mciteSetBstMidEndSepPunct{\mcitedefaultmidpunct}
{\mcitedefaultendpunct}{\mcitedefaultseppunct}\relax
\EndOfBibitem
\bibitem{Collins:1997sr}
J.~C. Collins, \ifthenelse{\boolean{articletitles}}{\emph{{Proof of
  factorization for diffractive hard scattering}},
  }{}\href{http://dx.doi.org/10.1103/PhysRevD.61.019902,
  10.1103/PhysRevD.57.3051}{Phys.\ Rev.\  \textbf{D57} (1998) 3051},
  \href{http://arxiv.org/abs/hep-ph/9709499}{{\normalfont\ttfamily
  arXiv:hep-ph/9709499}}, [Erratum: Phys. Rev.D61,019902(2000)]\relax
\mciteBstWouldAddEndPuncttrue
\mciteSetBstMidEndSepPunct{\mcitedefaultmidpunct}
{\mcitedefaultendpunct}{\mcitedefaultseppunct}\relax
\EndOfBibitem
\bibitem{Ingelman:1984ns}
G.~Ingelman and P.~E. Schlein, \ifthenelse{\boolean{articletitles}}{\emph{{Jet
  Structure in High Mass Diffractive Scattering}},
  }{}\href{http://dx.doi.org/10.1016/0370-2693(85)91181-5}{Phys.\ Lett.\
  \textbf{B152} (1985) 256}\relax
\mciteBstWouldAddEndPuncttrue
\mciteSetBstMidEndSepPunct{\mcitedefaultmidpunct}
{\mcitedefaultendpunct}{\mcitedefaultseppunct}\relax
\EndOfBibitem
\bibitem{datadifhera}
ZEUS, H1 Collaboration, F.~D. Aaron {\em et~al.},
  \ifthenelse{\boolean{articletitles}}{\emph{{Combined inclusive diffractive
  cross sections measured with forward proton spectrometers in deep inelastic
  $ep$ scattering at HERA}},
  }{}\href{http://dx.doi.org/10.1140/epjc/s10052-012-2175-y}{Eur.\ Phys.\ J.\
  \textbf{C72} (2012) 2175},
  \href{http://arxiv.org/abs/1207.4864}{{\normalfont\ttfamily
  arXiv:1207.4864}}\relax
\mciteBstWouldAddEndPuncttrue
\mciteSetBstMidEndSepPunct{\mcitedefaultmidpunct}
{\mcitedefaultendpunct}{\mcitedefaultseppunct}\relax
\EndOfBibitem
\bibitem{PhysRevD.47.101}
J.~D. Bjorken, \ifthenelse{\boolean{articletitles}}{\emph{Rapidity gaps and
  jets as a new-physics signature in very-high-energy hadron-hadron
  collisions}, }{}\href{http://dx.doi.org/10.1103/PhysRevD.47.101}{Phys.\ Rev.\
  D \textbf{47} (1993) 101}\relax
\mciteBstWouldAddEndPuncttrue
\mciteSetBstMidEndSepPunct{\mcitedefaultmidpunct}
{\mcitedefaultendpunct}{\mcitedefaultseppunct}\relax
\EndOfBibitem
\bibitem{kmr}
V.~A. Khoze, A.~D. Martin, and M.~G. Ryskin,
  \ifthenelse{\boolean{articletitles}}{\emph{{Elastic scattering and
  Diffractive dissociation in the light of LHC data}},
  }{}\href{http://dx.doi.org/10.1142/S0217751X1542004X}{Int.\ J.\ Mod.\ Phys.\
  \textbf{A30} (2015), no.~08 1542004},
  \href{http://arxiv.org/abs/1402.2778}{{\normalfont\ttfamily
  arXiv:1402.2778}}\relax
\mciteBstWouldAddEndPuncttrue
\mciteSetBstMidEndSepPunct{\mcitedefaultmidpunct}
{\mcitedefaultendpunct}{\mcitedefaultseppunct}\relax
\EndOfBibitem
\bibitem{Gotsman:2014pwa}
E.~Gotsman, E.~Levin, and U.~Maor,
  \ifthenelse{\boolean{articletitles}}{\emph{{A comprehensive model of soft
  interactions in the LHC era}},
  }{}\href{http://dx.doi.org/10.1142/S0217751X15420051}{Int.\ J.\ Mod.\ Phys.\
  \textbf{A30} (2015), no.~08 1542005},
  \href{http://arxiv.org/abs/1403.4531}{{\normalfont\ttfamily
  arXiv:1403.4531}}\relax
\mciteBstWouldAddEndPuncttrue
\mciteSetBstMidEndSepPunct{\mcitedefaultmidpunct}
{\mcitedefaultendpunct}{\mcitedefaultseppunct}\relax
\EndOfBibitem
\bibitem{roman}
B.~Kopeliovich, R.~Pasechnik, and I.~Potashnikova,
  \ifthenelse{\boolean{articletitles}}{\emph{{Hard hadronic diffraction is not
  hard}}, }{}\href{http://dx.doi.org/10.1142/S0218301316420015}{Int.\ J.\ Mod.\
  Phys.\  \textbf{E25} (2016), no.~07 1642001},
  \href{http://arxiv.org/abs/1603.08468}{{\normalfont\ttfamily
  arXiv:1603.08468}}\relax
\mciteBstWouldAddEndPuncttrue
\mciteSetBstMidEndSepPunct{\mcitedefaultmidpunct}
{\mcitedefaultendpunct}{\mcitedefaultseppunct}\relax
\EndOfBibitem
\bibitem{Khoze:2013dha}
V.~A. Khoze, A.~D. Martin, and M.~G. Ryskin,
  \ifthenelse{\boolean{articletitles}}{\emph{{Diffraction at the LHC}},
  }{}\href{http://dx.doi.org/10.1140/epjc/s10052-013-2503-x}{Eur.\ Phys.\ J.\
  \textbf{C73} (2013) 2503},
  \href{http://arxiv.org/abs/1306.2149}{{\normalfont\ttfamily
  arXiv:1306.2149}}\relax
\mciteBstWouldAddEndPuncttrue
\mciteSetBstMidEndSepPunct{\mcitedefaultmidpunct}
{\mcitedefaultendpunct}{\mcitedefaultseppunct}\relax
\EndOfBibitem
\bibitem{Gotsman:2011xc}
E.~Gotsman, E.~Levin, and U.~Maor,
  \ifthenelse{\boolean{articletitles}}{\emph{{Survival probability of large
  rapidity gaps in QCD and N=4 SYM motivated model}},
  }{}\href{http://dx.doi.org/10.1140/epjc/s10052-011-1685-3}{Eur.\ Phys.\ J.\
  \textbf{C71} (2011) 1685},
  \href{http://arxiv.org/abs/1101.5816}{{\normalfont\ttfamily
  arXiv:1101.5816}}\relax
\mciteBstWouldAddEndPuncttrue
\mciteSetBstMidEndSepPunct{\mcitedefaultmidpunct}
{\mcitedefaultendpunct}{\mcitedefaultseppunct}\relax
\EndOfBibitem
\bibitem{magno}
M.~B. Gay~Ducati, M.~M. Machado, and M.~V.~T. Machado,
  \ifthenelse{\boolean{articletitles}}{\emph{{Diffractive hadroproduction of
  $W^\pm$ and $Z^0$ bosons at high energies}},
  }{}\href{http://dx.doi.org/10.1103/PhysRevD.75.114013}{Phys.\ Rev.\
  \textbf{D75} (2007) 114013},
  \href{http://arxiv.org/abs/hep-ph/0703315}{{\normalfont\ttfamily
  arXiv:hep-ph/0703315}}\relax
\mciteBstWouldAddEndPuncttrue
\mciteSetBstMidEndSepPunct{\mcitedefaultmidpunct}
{\mcitedefaultendpunct}{\mcitedefaultseppunct}\relax
\EndOfBibitem
\bibitem{marta1}
M.~Luszczak, R.~Maciula, and A.~Szczurek,
  \ifthenelse{\boolean{articletitles}}{\emph{{Single- and central-diffractive
  production of open charm and bottom mesons at the LHC: theoretical
  predictions and experimental capabilities}},
  }{}\href{http://dx.doi.org/10.1103/PhysRevD.91.054024}{Phys.\ Rev.\
  \textbf{D91} (2015), no.~5 054024},
  \href{http://arxiv.org/abs/1412.3132}{{\normalfont\ttfamily
  arXiv:1412.3132}}\relax
\mciteBstWouldAddEndPuncttrue
\mciteSetBstMidEndSepPunct{\mcitedefaultmidpunct}
{\mcitedefaultendpunct}{\mcitedefaultseppunct}\relax
\EndOfBibitem
\bibitem{cristiano}
C.~Brenner~Mariotto and V.~P. Goncalves,
  \ifthenelse{\boolean{articletitles}}{\emph{{Diffractive photon production at
  the LHC}}, }{}\href{http://dx.doi.org/10.1103/PhysRevD.88.074023}{Phys.\
  Rev.\  \textbf{D88} (2013), no.~7 074023},
  \href{http://arxiv.org/abs/1309.2026}{{\normalfont\ttfamily
  arXiv:1309.2026}}\relax
\mciteBstWouldAddEndPuncttrue
\mciteSetBstMidEndSepPunct{\mcitedefaultmidpunct}
{\mcitedefaultendpunct}{\mcitedefaultseppunct}\relax
\EndOfBibitem
\bibitem{royon}
C.~Marquet, C.~Royon, M.~Saimpert, and D.~Werder,
  \ifthenelse{\boolean{articletitles}}{\emph{{Probing the Pomeron structure
  using dijets and $\gamma$+jet events at the LHC}},
  }{}\href{http://dx.doi.org/10.1103/PhysRevD.88.074029}{Phys.\ Rev.\
  \textbf{D88} (2013), no.~7 074029},
  \href{http://arxiv.org/abs/1306.4901}{{\normalfont\ttfamily
  arXiv:1306.4901}}\relax
\mciteBstWouldAddEndPuncttrue
\mciteSetBstMidEndSepPunct{\mcitedefaultmidpunct}
{\mcitedefaultendpunct}{\mcitedefaultseppunct}\relax
\EndOfBibitem
\bibitem{royon2}
A.~Chuinard, C.~Royon, and R.~Staszewski,
  \ifthenelse{\boolean{articletitles}}{\emph{{Testing Pomeron flavour symmetry
  with diffractive W charge asymmetry}},
  }{}\href{http://dx.doi.org/10.1007/JHEP04(2016)092}{JHEP \textbf{04} (2016)
  092}, \href{http://arxiv.org/abs/1510.04218}{{\normalfont\ttfamily
  arXiv:1510.04218}}\relax
\mciteBstWouldAddEndPuncttrue
\mciteSetBstMidEndSepPunct{\mcitedefaultmidpunct}
{\mcitedefaultendpunct}{\mcitedefaultseppunct}\relax
\EndOfBibitem
\bibitem{marta2}
M.~Luszczak, A.~Szczurek, and C.~Royon,
  \ifthenelse{\boolean{articletitles}}{\emph{{$W^+ W^-$ pair production in
  proton-proton collisions: small missing terms}},
  }{}\href{http://dx.doi.org/10.1007/JHEP02(2015)098}{JHEP \textbf{02} (2015)
  098}, \href{http://arxiv.org/abs/1409.1803}{{\normalfont\ttfamily
  arXiv:1409.1803}}\relax
\mciteBstWouldAddEndPuncttrue
\mciteSetBstMidEndSepPunct{\mcitedefaultmidpunct}
{\mcitedefaultendpunct}{\mcitedefaultseppunct}\relax
\EndOfBibitem
\bibitem{cristiano2}
C.~Brenner~Mariotto and V.~P. Goncalves,
  \ifthenelse{\boolean{articletitles}}{\emph{{Double $J/\psi$ production in
  central diffractive processes at the LHC}},
  }{}\href{http://dx.doi.org/10.1103/PhysRevD.91.114002}{Phys.\ Rev.\
  \textbf{D91} (2015), no.~11 114002},
  \href{http://arxiv.org/abs/1502.02612}{{\normalfont\ttfamily
  arXiv:1502.02612}}\relax
\mciteBstWouldAddEndPuncttrue
\mciteSetBstMidEndSepPunct{\mcitedefaultmidpunct}
{\mcitedefaultendpunct}{\mcitedefaultseppunct}\relax
\EndOfBibitem
\bibitem{nos_prd}
V.~P. Goncalves, C.~Potterat, and M.~S. Rangel,
  \ifthenelse{\boolean{articletitles}}{\emph{{Bottom production in Photon and
  Pomeron -- induced interactions at the LHC}},
  }{}\href{http://dx.doi.org/10.1103/PhysRevD.93.034038}{Phys.\ Rev.\
  \textbf{D93} (2016), no.~3 034038},
  \href{http://arxiv.org/abs/1511.07688}{{\normalfont\ttfamily
  arXiv:1511.07688}}\relax
\mciteBstWouldAddEndPuncttrue
\mciteSetBstMidEndSepPunct{\mcitedefaultmidpunct}
{\mcitedefaultendpunct}{\mcitedefaultseppunct}\relax
\EndOfBibitem
\bibitem{marquet}
C.~Marquet {\em et~al.},
  \ifthenelse{\boolean{articletitles}}{\emph{{Diffractive di-jet production at
  the LHC with a Reggeon contribution}},
  }{}\href{http://arxiv.org/abs/1608.05674}{{\normalfont\ttfamily
  arXiv:1608.05674}}\relax
\mciteBstWouldAddEndPuncttrue
\mciteSetBstMidEndSepPunct{\mcitedefaultmidpunct}
{\mcitedefaultendpunct}{\mcitedefaultseppunct}\relax
\EndOfBibitem
\bibitem{marta3}
M.~Luszczak, R.~Maciula, A.~Szczurek, and M.~Trzebinski,
  \ifthenelse{\boolean{articletitles}}{\emph{{Single-diffractive production of
  charmed mesons at the LHC within the $k_t$-factorization approach}},
  }{}\href{http://arxiv.org/abs/1606.06528}{{\normalfont\ttfamily
  arXiv:1606.06528}}\relax
\mciteBstWouldAddEndPuncttrue
\mciteSetBstMidEndSepPunct{\mcitedefaultmidpunct}
{\mcitedefaultendpunct}{\mcitedefaultseppunct}\relax
\EndOfBibitem
\bibitem{Rasmussen:2015qgr}
C.~O. Rasmussen and T.~Sj{\"o}strand,
  \ifthenelse{\boolean{articletitles}}{\emph{{Hard Diffraction with Dynamic Gap
  Survival}}, }{}\href{http://dx.doi.org/10.1007/JHEP02(2016)142}{JHEP
  \textbf{02} (2016) 142},
  \href{http://arxiv.org/abs/1512.05525}{{\normalfont\ttfamily
  arXiv:1512.05525}}\relax
\mciteBstWouldAddEndPuncttrue
\mciteSetBstMidEndSepPunct{\mcitedefaultmidpunct}
{\mcitedefaultendpunct}{\mcitedefaultseppunct}\relax
\EndOfBibitem
\bibitem{Sjostrand:2014zea}
T.~Sj{\"o}strand {\em et~al.}, \ifthenelse{\boolean{articletitles}}{\emph{{An
  Introduction to PYTHIA 8.2}},
  }{}\href{http://dx.doi.org/10.1016/j.cpc.2015.01.024}{Comput.\ Phys.\
  Commun.\  \textbf{191} (2015) 159},
  \href{http://arxiv.org/abs/1410.3012}{{\normalfont\ttfamily
  arXiv:1410.3012}}\relax
\mciteBstWouldAddEndPuncttrue
\mciteSetBstMidEndSepPunct{\mcitedefaultmidpunct}
{\mcitedefaultendpunct}{\mcitedefaultseppunct}\relax
\EndOfBibitem
\bibitem{review_FPWG}
LHC Forward Physics Working Group, e.~Royon, C.\ {\em et~al.},
  \ifthenelse{\boolean{articletitles}}{\emph{{LHC Forward Physics}},
  }{}CERN-PH-LPCC-2015-001, SLAC-PUB-16364, DESY-15-167 (2015)\relax
\mciteBstWouldAddEndPuncttrue
\mciteSetBstMidEndSepPunct{\mcitedefaultmidpunct}
{\mcitedefaultendpunct}{\mcitedefaultseppunct}\relax
\EndOfBibitem
\bibitem{LHCb-DP-2014-002}
LHCb Collaboration, R.~Aaij {\em et~al.},
  \ifthenelse{\boolean{articletitles}}{\emph{{LHCb detector performance}},
  }{}\href{http://dx.doi.org/10.1142/S0217751X15300227}{Int.\ J.\ Mod.\ Phys.\
  \textbf{A30} (2015) 1530022},
  \href{http://arxiv.org/abs/1412.6352}{{\normalfont\ttfamily
  arXiv:1412.6352}}\relax
\mciteBstWouldAddEndPuncttrue
\mciteSetBstMidEndSepPunct{\mcitedefaultmidpunct}
{\mcitedefaultendpunct}{\mcitedefaultseppunct}\relax
\EndOfBibitem
\bibitem{LHCb-PAPER-2016-021}
LHCb Collaboration, R.~Aaij {\em et~al.},
  \ifthenelse{\boolean{articletitles}}{\emph{{Measurement of the forward $Z$
  boson production cross-section in $pp$ collisions at $\sqrt{s}=13$ TeV}}, }{}
  {LHCb-PAPER-2016-021}, {in preparation}\relax
\mciteBstWouldAddEndPuncttrue
\mciteSetBstMidEndSepPunct{\mcitedefaultmidpunct}
{\mcitedefaultendpunct}{\mcitedefaultseppunct}\relax
\EndOfBibitem
\bibitem{LHCb-PAPER-2014-033}
LHCb Collaboration, R.~Aaij {\em et~al.},
  \ifthenelse{\boolean{articletitles}}{\emph{{Measurement of the forward $W$
  boson production cross-section in $pp$ collisions at $\sqrt{s}=7$ TeV}},
  }{}\href{http://dx.doi.org/10.1007/JHEP12(2014)079}{JHEP \textbf{12} (2014)
  079}, \href{http://arxiv.org/abs/1408.4354}{{\normalfont\ttfamily
  arXiv:1408.4354}}\relax
\mciteBstWouldAddEndPuncttrue
\mciteSetBstMidEndSepPunct{\mcitedefaultmidpunct}
{\mcitedefaultendpunct}{\mcitedefaultseppunct}\relax
\EndOfBibitem
\bibitem{LHCb-PAPER-2015-001}
LHCb Collaboration, R.~Aaij {\em et~al.},
  \ifthenelse{\boolean{articletitles}}{\emph{{Measurement of the forward $Z$
  boson cross-section in $pp$ collisions at $\sqrt{s}=7$ TeV}},
  }{}\href{http://dx.doi.org/10.1007/JHEP08(2015)039}{JHEP \textbf{08} (2015)
  039}, \href{http://arxiv.org/abs/1505.07024}{{\normalfont\ttfamily
  arXiv:1505.07024}}\relax
\mciteBstWouldAddEndPuncttrue
\mciteSetBstMidEndSepPunct{\mcitedefaultmidpunct}
{\mcitedefaultendpunct}{\mcitedefaultseppunct}\relax
\EndOfBibitem
\bibitem{LHCb-PAPER-2015-049}
LHCb Collaboration, R.~Aaij {\em et~al.},
  \ifthenelse{\boolean{articletitles}}{\emph{{Measurement of forward $W$ and
  $Z$ boson production in $pp$ collisions at $\sqrt{s}=8$ TeV }},
  }{}\href{http://dx.doi.org/10.1007/JHEP01(2016)155}{JHEP \textbf{01} (2015)
  155}, \href{http://arxiv.org/abs/1511.08039}{{\normalfont\ttfamily
  arXiv:1511.08039}}\relax
\mciteBstWouldAddEndPuncttrue
\mciteSetBstMidEndSepPunct{\mcitedefaultmidpunct}
{\mcitedefaultendpunct}{\mcitedefaultseppunct}\relax
\EndOfBibitem
\bibitem{LHCb-PAPER-2013-058}
LHCb Collaboration, R.~Aaij {\em et~al.},
  \ifthenelse{\boolean{articletitles}}{\emph{{Study of forward $Z$+jet
  production in $pp$ collisions at $\sqrt{s} = 7$ TeV}},
  }{}\href{http://dx.doi.org/10.1007/JHEP01(2014)033}{JHEP \textbf{01} (2014)
  033}, \href{http://arxiv.org/abs/1310.8197}{{\normalfont\ttfamily
  arXiv:1310.8197}}\relax
\mciteBstWouldAddEndPuncttrue
\mciteSetBstMidEndSepPunct{\mcitedefaultmidpunct}
{\mcitedefaultendpunct}{\mcitedefaultseppunct}\relax
\EndOfBibitem
\bibitem{LHCb-PAPER-2014-055}
LHCb Collaboration, R.~Aaij {\em et~al.},
  \ifthenelse{\boolean{articletitles}}{\emph{{Measurement of the $Z$+$b$-jet
  cross-section in $pp$ collisions at $\sqrt{s} = 7$ TeV in the forward
  region}}, }{}\href{http://dx.doi.org/10.1007/JHEP01(2015)064}{JHEP
  \textbf{01} (2015) 064},
  \href{http://arxiv.org/abs/1411.1264}{{\normalfont\ttfamily
  arXiv:1411.1264}}\relax
\mciteBstWouldAddEndPuncttrue
\mciteSetBstMidEndSepPunct{\mcitedefaultmidpunct}
{\mcitedefaultendpunct}{\mcitedefaultseppunct}\relax
\EndOfBibitem
\bibitem{LHCb-PAPER-2015-021}
LHCb Collaboration, R.~Aaij {\em et~al.},
  \ifthenelse{\boolean{articletitles}}{\emph{{Study of $W$ boson production in
  association with beauty and charm}},
  }{}\href{http://dx.doi.org/10.1103/PhysRevD.92.052001}{Phys.\ Rev.\
  \textbf{D92} (2015) 052012},
  \href{http://arxiv.org/abs/1505.04051}{{\normalfont\ttfamily
  arXiv:1505.04051}}\relax
\mciteBstWouldAddEndPuncttrue
\mciteSetBstMidEndSepPunct{\mcitedefaultmidpunct}
{\mcitedefaultendpunct}{\mcitedefaultseppunct}\relax
\EndOfBibitem
\bibitem{LHCb-PAPER-2016-011}
LHCb Collaboration, R.~Aaij {\em et~al.},
  \ifthenelse{\boolean{articletitles}}{\emph{{Measurement of forward $W$ and
  $Z$ boson production in association with jets in proton-proton collisions at
  $\sqrt{s}=8$\,TeV}}, }{}\href{http://dx.doi.org/10.1007/JHEP05(2016)131}{JHEP
  \textbf{05} (2016) 131},
  \href{http://arxiv.org/abs/1605.00951}{{\normalfont\ttfamily
  arXiv:1605.00951}}\relax
\mciteBstWouldAddEndPuncttrue
\mciteSetBstMidEndSepPunct{\mcitedefaultmidpunct}
{\mcitedefaultendpunct}{\mcitedefaultseppunct}\relax
\EndOfBibitem
\bibitem{lhcbYcep}
LHCb Collaboration, R.~Aaij {\em et~al.},
  \ifthenelse{\boolean{articletitles}}{\emph{{Measurement of the exclusive Υ
  production cross-section in pp collisions at $ \sqrt{s}=7 $ TeV and 8 TeV}},
  }{}\href{http://dx.doi.org/10.1007/JHEP09(2015)084}{JHEP \textbf{09} (2015)
  084}, \href{http://arxiv.org/abs/1505.08139}{{\normalfont\ttfamily
  arXiv:1505.08139}}\relax
\mciteBstWouldAddEndPuncttrue
\mciteSetBstMidEndSepPunct{\mcitedefaultmidpunct}
{\mcitedefaultendpunct}{\mcitedefaultseppunct}\relax
\EndOfBibitem
\bibitem{lhcbJPSIcep}
LHCb Collaboration, R.~Aaij {\em et~al.},
  \ifthenelse{\boolean{articletitles}}{\emph{{Updated measurements of exclusive
  $J/\psi$ and $\psi$(2S) production cross-sections in pp collisions at
  $\sqrt{s}=7$ TeV}},
  }{}\href{http://dx.doi.org/10.1088/0954-3899/41/5/055002}{J.\ Phys.\
  \textbf{G41} (2014) 055002},
  \href{http://arxiv.org/abs/1401.3288}{{\normalfont\ttfamily
  arXiv:1401.3288}}\relax
\mciteBstWouldAddEndPuncttrue
\mciteSetBstMidEndSepPunct{\mcitedefaultmidpunct}
{\mcitedefaultendpunct}{\mcitedefaultseppunct}\relax
\EndOfBibitem
\bibitem{lhcbDDcep}
LHCb Collaboration, R.~Aaij {\em et~al.},
  \ifthenelse{\boolean{articletitles}}{\emph{{Observation of charmonium pairs
  produced exclusively in $pp$ collisions}},
  }{}\href{http://dx.doi.org/10.1088/0954-3899/41/11/115002}{J.\ Phys.\
  \textbf{G41} (2014), no.~11 115002},
  \href{http://arxiv.org/abs/1407.5973}{{\normalfont\ttfamily
  arXiv:1407.5973}}\relax
\mciteBstWouldAddEndPuncttrue
\mciteSetBstMidEndSepPunct{\mcitedefaultmidpunct}
{\mcitedefaultendpunct}{\mcitedefaultseppunct}\relax
\EndOfBibitem
\bibitem{LHCb-DP-2016-003}
K.~C. Akiba {\em et~al.}, \ifthenelse{\boolean{articletitles}}{\emph{{The
  Herschel detector: high rapidity shower counters for LHCb}},
  }{}LHCb-DP-2016-003\relax
\mciteBstWouldAddEndPuncttrue
\mciteSetBstMidEndSepPunct{\mcitedefaultmidpunct}
{\mcitedefaultendpunct}{\mcitedefaultseppunct}\relax
\EndOfBibitem
\bibitem{Jpsi13tev}
LHCb Collaboration, R.~Aaij {\em et~al.},
  \ifthenelse{\boolean{articletitles}}{\emph{{Central exclusive production of
  $J/\psi$ and $\psi(2S)$ mesons in pp collisions at $\sqrt{s}=13$ TeV}},
  }{}LHCb-CONF-2016-007, CERN-LHCb-CONF-2016-007 (2016)\relax
\mciteBstWouldAddEndPuncttrue
\mciteSetBstMidEndSepPunct{\mcitedefaultmidpunct}
{\mcitedefaultendpunct}{\mcitedefaultseppunct}\relax
\EndOfBibitem
\bibitem{Aktas:2006hy}
H1 Collaboration, A.~Aktas {\em et~al.},
  \ifthenelse{\boolean{articletitles}}{\emph{{Measurement and QCD analysis of
  the diffractive deep-inelastic scattering cross-section at HERA}},
  }{}\href{http://dx.doi.org/10.1140/epjc/s10052-006-0035-3}{Eur.\ Phys.\ J.\
  \textbf{C48} (2006) 715},
  \href{http://arxiv.org/abs/hep-ex/0606004}{{\normalfont\ttfamily
  arXiv:hep-ex/0606004}}\relax
\mciteBstWouldAddEndPuncttrue
\mciteSetBstMidEndSepPunct{\mcitedefaultmidpunct}
{\mcitedefaultendpunct}{\mcitedefaultseppunct}\relax
\EndOfBibitem
\bibitem{Aktas:2006hx}
H1 Collaboration, A.~Aktas {\em et~al.},
  \ifthenelse{\boolean{articletitles}}{\emph{{Diffractive deep-inelastic
  scattering with a leading proton at HERA}},
  }{}\href{http://dx.doi.org/10.1140/epjc/s10052-006-0046-0}{Eur.\ Phys.\ J.\
  \textbf{C48} (2006) 749},
  \href{http://arxiv.org/abs/hep-ex/0606003}{{\normalfont\ttfamily
  arXiv:hep-ex/0606003}}\relax
\mciteBstWouldAddEndPuncttrue
\mciteSetBstMidEndSepPunct{\mcitedefaultmidpunct}
{\mcitedefaultendpunct}{\mcitedefaultseppunct}\relax
\EndOfBibitem
\bibitem{ct10}
H.-L. Lai {\em et~al.}, \ifthenelse{\boolean{articletitles}}{\emph{{Parton
  Distributions for Event Generators}},
  }{}\href{http://dx.doi.org/10.1007/JHEP04(2010)035}{JHEP \textbf{04} (2010)
  035}, \href{http://arxiv.org/abs/0910.4183}{{\normalfont\ttfamily
  arXiv:0910.4183}}\relax
\mciteBstWouldAddEndPuncttrue
\mciteSetBstMidEndSepPunct{\mcitedefaultmidpunct}
{\mcitedefaultendpunct}{\mcitedefaultseppunct}\relax
\EndOfBibitem
\bibitem{Edin:1995gi}
A.~Edin, G.~Ingelman, and J.~Rathsman,
  \ifthenelse{\boolean{articletitles}}{\emph{{Soft color interactions as the
  origin of rapidity gaps in DIS}},
  }{}\href{http://dx.doi.org/10.1016/0370-2693(95)01391-1}{Phys.\ Lett.\
  \textbf{B366} (1996) 371},
  \href{http://arxiv.org/abs/hep-ph/9508386}{{\normalfont\ttfamily
  arXiv:hep-ph/9508386}}\relax
\mciteBstWouldAddEndPuncttrue
\mciteSetBstMidEndSepPunct{\mcitedefaultmidpunct}
{\mcitedefaultendpunct}{\mcitedefaultseppunct}\relax
\EndOfBibitem
\bibitem{Buchmuller:1995qa}
W.~Buchmuller and A.~Hebecker, \ifthenelse{\boolean{articletitles}}{\emph{{A
  Parton model for diffractive processes in deep inelastic scattering}},
  }{}\href{http://dx.doi.org/10.1016/0370-2693(95)00721-V}{Phys.\ Lett.\
  \textbf{B355} (1995) 573},
  \href{http://arxiv.org/abs/hep-ph/9504374}{{\normalfont\ttfamily
  arXiv:hep-ph/9504374}}\relax
\mciteBstWouldAddEndPuncttrue
\mciteSetBstMidEndSepPunct{\mcitedefaultmidpunct}
{\mcitedefaultendpunct}{\mcitedefaultseppunct}\relax
\EndOfBibitem
\bibitem{Brodsky:2002ue}
S.~J. Brodsky {\em et~al.},
  \ifthenelse{\boolean{articletitles}}{\emph{{Structure functions are not
  parton probabilities}},
  }{}\href{http://dx.doi.org/10.1103/PhysRevD.65.114025}{Phys.\ Rev.\
  \textbf{D65} (2002) 114025},
  \href{http://arxiv.org/abs/hep-ph/0104291}{{\normalfont\ttfamily
  arXiv:hep-ph/0104291}}\relax
\mciteBstWouldAddEndPuncttrue
\mciteSetBstMidEndSepPunct{\mcitedefaultmidpunct}
{\mcitedefaultendpunct}{\mcitedefaultseppunct}\relax
\EndOfBibitem
\bibitem{Brodsky:2004hi}
S.~J. Brodsky, R.~Enberg, P.~Hoyer, and G.~Ingelman,
  \ifthenelse{\boolean{articletitles}}{\emph{{Hard diffraction from parton
  rescattering in QCD}},
  }{}\href{http://dx.doi.org/10.1103/PhysRevD.71.074020}{Phys.\ Rev.\
  \textbf{D71} (2005) 074020},
  \href{http://arxiv.org/abs/hep-ph/0409119}{{\normalfont\ttfamily
  arXiv:hep-ph/0409119}}\relax
\mciteBstWouldAddEndPuncttrue
\mciteSetBstMidEndSepPunct{\mcitedefaultmidpunct}
{\mcitedefaultendpunct}{\mcitedefaultseppunct}\relax
\EndOfBibitem
\bibitem{Pasechnik:2010zs}
R.~Pasechnik, R.~Enberg, and G.~Ingelman,
  \ifthenelse{\boolean{articletitles}}{\emph{{QCD rescattering mechanism for
  diffractive deep inelastic scattering}},
  }{}\href{http://dx.doi.org/10.1103/PhysRevD.82.054036}{Phys.\ Rev.\
  \textbf{D82} (2010) 054036},
  \href{http://arxiv.org/abs/1005.3399}{{\normalfont\ttfamily
  arXiv:1005.3399}}\relax
\mciteBstWouldAddEndPuncttrue
\mciteSetBstMidEndSepPunct{\mcitedefaultmidpunct}
{\mcitedefaultendpunct}{\mcitedefaultseppunct}\relax
\EndOfBibitem
\bibitem{Ingelman:2015qrt}
G.~Ingelman, R.~Pasechnik, and D.~Werder,
  \ifthenelse{\boolean{articletitles}}{\emph{{Dynamic color screening in
  diffractive deep inelastic scattering}},
  }{}\href{http://dx.doi.org/10.1103/PhysRevD.93.094016}{Phys.\ Rev.\
  \textbf{D93} (2016), no.~9 094016},
  \href{http://arxiv.org/abs/1511.06317}{{\normalfont\ttfamily
  arXiv:1511.06317}}\relax
\mciteBstWouldAddEndPuncttrue
\mciteSetBstMidEndSepPunct{\mcitedefaultmidpunct}
{\mcitedefaultendpunct}{\mcitedefaultseppunct}\relax
\EndOfBibitem
\bibitem{antikt}
M.~Cacciari, G.~P. Salam, and G.~Soyez,
  \ifthenelse{\boolean{articletitles}}{\emph{{The Anti-k(t) jet clustering
  algorithm}}, }{}\href{http://dx.doi.org/10.1088/1126-6708/2008/04/063}{JHEP
  \textbf{04} (2008) 063},
  \href{http://arxiv.org/abs/0802.1189}{{\normalfont\ttfamily
  arXiv:0802.1189}}\relax
\mciteBstWouldAddEndPuncttrue
\mciteSetBstMidEndSepPunct{\mcitedefaultmidpunct}
{\mcitedefaultendpunct}{\mcitedefaultseppunct}\relax
\EndOfBibitem
\bibitem{fastjet}
M.~Cacciari, G.~P. Salam, and G.~Soyez,
  \ifthenelse{\boolean{articletitles}}{\emph{{FastJet user manual}},
  }{}\href{http://arxiv.org/abs/1111.6097}{{\normalfont\ttfamily
  arXiv:1111.6097}}\relax
\mciteBstWouldAddEndPuncttrue
\mciteSetBstMidEndSepPunct{\mcitedefaultmidpunct}
{\mcitedefaultendpunct}{\mcitedefaultseppunct}\relax
\EndOfBibitem
\end{mcitethebibliography}

\newpage


\end{document}